\documentclass[aps,prc,twocolumn, superscriptaddress,showkeys,nofootinbib]{revtex4-1}  

\usepackage[utf8]{inputenc}

\usepackage{graphicx,color}
\usepackage{amsmath,amssymb,amsfonts}
\usepackage{hyperref}

\usepackage{xspace}	

\def  \etas       {\mbox{$\eta / \textit{s}$ }\xspace}

\def \cnm2   {\mbox{$\mathrm{C}_{nm}\lbrace 2 \rbrace$ }\xspace}
\def \cnm4   {\mbox{$\mathrm{C}_{nm}\lbrace 4 \rbrace$ }\xspace}

\def \sc23  {\mbox{$\mathrm{SC}(2,3)$ }\xspace}
\def \sc24  {\mbox{$\mathrm{SC}(2,4)$ }\xspace}

\def \nsc23 {\mbox{$\mathrm{NSC}(2,3)$}\xspace}
\def \nsc24 {\mbox{$\mathrm{NSC}(2,4)$}\xspace}

\newcommand{\psirp}{\Psi^{\rm  RP}}

\begin{document}
\title{Characterizing the initial and final state effects of relativistic nuclear collisions}
\medskip
\author{Niseem~Magdy} 
\email{niseemm@gmail.com}
\affiliation{Department of Chemistry, State University of New York, Stony Brook, New York 11794, USA}

\begin{abstract}
The  Multi-Phase Transport model (AMPT) is used to study the final state effects on the Symmetric Correlations (SC), Asymmetric Correlations (ASC), Normalized Symmetric Correlations (NSC), and Normalized Asymmetric Correlations (NASC) in Au+Au collisions at 200~GeV. The correlators' sensitivity to non-flow effects associated with long- and short-range non-flow correlations are also discussed using the HIJING model. The results indicate that SC, ASC, NSC, and NASC can give accompanying constraints for initial and final state effects. In addition, conducting further detailed experimental measurements spanning a broad range of collision systems and beam energies will serve as an additional constraint for the theoretical models' calculations.
\end{abstract}
\maketitle
\section{Introduction}
Investigations at Relativistic Heavy Ion Collider (RHIC) and  Large Hadron Collider (LHC) seek to explore the properties of the deconfined nuclear matter called quark-gluon plasma (QGP)~\cite{Shuryak:1978ij,Shuryak:1980tp,Muller:2012zq}. 
In relativistic nuclear collisions, the created QGP will experience a hydrodynamic evolution. The event-by-event fluctuating participant distribution defines the initial geometry of the collision.
A central aim of prior and current experimental and theoretical investigations of quark-gluon plasma is to understand its transport properties, such as specific shear viscosity (\etas)~\cite{Shuryak:2003xe,Romatschke:2007mq,Luzum:2008cw,Bozek:2009dw,ALICE:2019zfl,Parkkila:2021yha,ALICE:2020sup,Adam:2020ymj}. 

Anisotropic flow measurements continue to be a valuable route to \etas estimation because they reflect the viscous hydrodynamic response to the anisotropy of the initial state energy density~\cite{Bozek:2020drh,Alver:2008zza,Giacalone:2020byk,Schenke:2014tga,Staig:2010pn,Heinz:2001xi,Hirano:2005xf,Huovinen:2001cy,Hirano:2002ds,Romatschke:2007mq,Luzum:2011mm,Song:2010mg,Qian:2016fpi,Magdy:2017ohf,Magdy:2017kji,Bilandzic:2021voo,Schenke:2011tv,Teaney:2012ke,Gardim:2012yp,Lacey:2013eia}, which is illustrated by the complex eccentricity vectors $\mathcal{E}_{n}$~\cite{Alver:2010dn,Petersen:2010cw,Lacey:2010hw,Teaney:2010vd,Qiu:2011iv}:
\begin{eqnarray}
\mathcal{E}_{n} & \equiv &  \varepsilon_{n} e^{i {\textit{n}} \Phi_{n} }  \\  \nonumber
&\equiv  &
  - \frac{\int dx^{'}\,dy^{'}\,\textit{r}^{n}\,e^{i {\textit{n}} \phi}\, \textit{E}(r,\phi)}
           {\int dx^{'}\,dy^{'}\,\textit{r}^{n}\,\textit{E}(r,\phi)}, ~(\textit{n} ~>~ 1),
\label{epsdef1}
\end{eqnarray}
where  $\varepsilon_{n}$ and $\mathrm{\Phi_{n}}$ are the magnitude and angular direction of the n$^{\rm th}$ eccentricity vector. The $r$ is the radial coordinate, $\phi$ is the spatial azimuthal angle, and ${\textit{E}}(r,\phi)$ is the initial energy density profile~\cite{Teaney:2010vd,Bhalerao:2014xra,Yan:2015jma}. 

In heavy ion collisions, the medium experiences a pressure gradient. Such gradient transforms the initial state's spatial anisotropies into final state momentum anisotropies. Consequently, the azimuthal distribution of the particles produced in the collisions can be analyzed with a Fourier expansion as~\cite{Voloshin:1994mz,Poskanzer:1998yz}:
\begin{eqnarray}
\dfrac{dN}{d\varphi} &\propto & 1 + \sum_{n=1}^{\infty} 2 v_{n} \cos\left[  n(\varphi -\psirp) \right] ,
\label{equ:Fourier_expansion}   
\end{eqnarray}
where $\varphi$ defines the azimuthal angle of the particle, $v_n$ is the n$^{th}$-order flow harmonic, and $\psirp$ is the reaction plane. 

Although the n$^{th}$-order flow harmonics are linear related to the same order initial state anisotropies $\varepsilon_{{{n}}}$, the higher-order flow harmonics, n$>$3, in addition, includes a non-linear response to the lower-order eccentricities $\varepsilon_{{{n=2, 3}}}$~\cite{Song:2010mg, Niemi:2012aj,Gardim:2014tya,Fu:2015wba,Holopainen:2010gz,Qin:2010pf,Qiu:2011iv,Gale:2012rq,Liu:2018hjh,Teaney:2012ke,Bhalerao:2014xra,Yan:2015jma,Gardim:2011xv}. Such a relationship encodes the medium response (i.e., it's sensitive to $\eta/s$).
An accurate estimation of $\eta/s$ requires powerful model constraints for initial state eccentricities and their fluctuations and correlations across a broad range of beam energies and collision systems~\cite{Schenke:2019ruo,Alba:2017hhe}. These model constraints are attainable by providing a comprehensive set of data model comparisons of the Symmetric Correlations (SC), Asymmetric Correlations (ASC), Normalized Symmetric Correlations (NSC), and Normalized Asymmetric Correlations (NASC). These correlations have been studied in Pb--Pb collisions at Large Hadron Collider (LHC) energies using hydrodynamic models. Using different analysis methods and varied initial conditions, the six- and eight-particles SC and NSC were investigated in Refs~\cite{Mordasini:2019hut,Moravcova:2020wnf,Li:2021nas,Bilandzic:2021rgb}. In addition, the NASC, the event plane angular correlations, for harmonics 2--5 were studied in Refs~\cite{Bilandzic:2021rgb,Bilandzic:2020csw,Taghavi:2020gcy}.

In this work, the prior investigations of SC, ASC, NSC, and NASC~\cite{Taghavi:2020gcy,Bilandzic:2021rgb,Bilandzic:2020csw,Mordasini:2019hut,Li:2021nas,Moravcova:2020wnf} have been extend to the Au+Au at RHIC top energy and flow harmonics $v_1$--$v_6$. In addition,  I systematically investigate the influence of the final state effects (i.e., medium properties \etas, hadronization, and hadronic transport) using the AMPT model~\cite{Lin:2004en} on the SC, ASC, NSC, and NASC. The presented AMPT calculations are also compared with the experimental data (when available). 
The data model comparisons will reflect the SC, ASC, NSC, and NASC potential to constrain the heavy ion collisions' initial and final state effects given by the AMPT model. Therefore, providing beam-energy and system-size measurements of the SC, ASC, NSC, and NASC can give simultaneous constraints on the initial and final state effects and help decrement between various theoretical models.

The current work also consider the effects of long-range non-flow (e.g., jets in a dijet event) and short-range non-flow (e.g., resonance decays, Bose-Einstein correlation, and fragments of individual jets) on the presented quantities using the HIJING model~\cite{Wang:1991hta,Gyulassy:1994ew} that containing only non-flow correlations. 
Non-flow effects commonly involve a few particles from one or more regions in  pseudorapidity ($\eta$). The non-flow effects are usually reduced by correlating (I) multi-particles (i.e., 5- and 6-particles) from the same pseudorapidity and (II) particles from two or more subevents divided in pseudorapidity~\cite{Jia:2017hbm}. Therefore, a careful analysis of the non-flow effects on SC and ASC is required before interpreting the experimental measurements. 

The paper is organized as follows. Section~\ref{sec:2} summarizes the theoretical models used to investigate the SC, ASC, NSC, and NASC and the details of the analysis method employed. The results from my studies are presented in Sec.~\ref{sec:3} followed by a summary and outlook in Sec~\ref{sec:4}.

\section{Methodology} \label{sec:2}
\subsection{Models}
The current analysis is conducted with events simulated by the AMPT (v2.26t9b)~\cite{Lin:2004en}  and HIJING (v1.411)~\cite{Wang:1991hta,Gyulassy:1994ew} models, for Au+Au collisions at  $\sqrt{s_{NN}}$ = 200~GeV. In the used models, charged particles with $0.2 < p_T < 2.0$ GeV/$c$, and $|\eta| < 1.0$ were chosen for analysis. The HIJING model highlights the non-flow effects, while the AMPT model investigates the flow correlators' dependence on final state effects.

{\color{black}
The HIJING model~\cite{Wang:1991hta,Gyulassy:1994ew} is a Monte Carlo event generator that simulates the parton and particle production in high-energy heavy ion collisions. Using the Glauber geometry, the HIJING model can simulate heavy-ion collisions via binary nucleon-nucleon collisions. The model is used to study jet and mini-jet production and associated particle production in high-energy p+p, p+A, and A+A collisions. 
In addition, the HIJING model employs PYTHIA~\cite{Sjostrand:1986hx} for developing kinematic variables of scattered partons for each hard- or semihard-interaction and Lund string fragmentation~\cite{Andersson:1983ia} for hadronization processes. 
}


The AMPT model is broadly used to investigate  heavy-ion collisions physics~\cite{Lin:2004en,Ma:2016fve,Ma:2013gga,Ma:2013uqa,Bzdak:2014dia,Nie:2018xog,Haque:2019vgi,Zhao:2019kyk,Bhaduri:2010wi,Nasim:2010hw,Xu:2010du,Magdy:2020bhd,Guo:2019joy,Magdy:2020gxf,Magdy:2022cvt}. 
The model contains several fundamental components: (i) an initial partonic condition given by the HIJING model~\cite{Wang:1991hta,Gyulassy:1994ew}. 
The parameters of the HIJING model that is used in the  Lund string fragmentation function~\cite{Ferreres-Sole:2018vgo}
($f(z) \propto z^{-1} (1-z)^a\exp (-b~m_{\perp}^2/z)$) are; $a=0.55$ and $b=0.15$ GeV$^{-2}$. {\color{black} Such parameters are given in Ref~\cite{Xu:2011fi}}.
Note that $z$ denotes the light-cone momentum fraction of the yielded hadron of transverse mass $m_\perp$ around that of the fragmenting string. (ii) partonic scattering with a cross-section,
\begin{eqnarray}\label{eq:21}
\sigma_{pp} &=& \dfrac{9 \pi \alpha^{2}_{s}}{2 \mu^{2}},
\end{eqnarray}
where $\mu$ stands for the screening mass, and $\alpha_{s}$  denotes the QCD coupling constant. They typically represent the expansion dynamics of the A--A collision~\cite{Zhang:1997ej}.
Within the AMPT framework, the initial value of \etas (evaluated at the beginning of heavy-ion collision) can be related to $\sigma_{pp}$ see Refs~\cite{Xu:2011fi,Nasim:2016rfv,Solanki:2012ne}.
 (iii) the hadronization process via coalescence followed by the hadronic interactions~\cite{Li:1995pra}. 
In the current work, Au+Au collisions at $\sqrt{s_{\rm NN}}=$ 200~GeV, were simulated with AMPT version ampt-v2.26t9b for a fixed value of $\alpha_{s}$ = 0.47, but varying $\mu$ as given in Tab.~\ref{tab:1}. {\color{black} The model parameters ranges used in this work are inspired by the studies in Refs~\cite{Xu:2011fi,Nasim:2016rfv,Magdy:2021sba,Magdy:2020fma}}.

\begin{table}[h!]
\begin{center}
\caption{The summary of the AMPT sets used in this work.\label{tab:1}}
 \begin{tabular}{|c|c|c|c|}
 \hline 
 AMPT-set      &  $\mu$~($fm^{-1}$) &   $\sigma_{pp}$   &  \etas  \\
  \hline
  Set-1        &         3.35       &       2.8         &   0.2    \\
  \hline
  Set-2        &         4.20       &       1.8         &   0.3    \\
 \hline
\end{tabular} 
\end{center}
\end{table}

The results presented in this work were obtained for minimum bias Au+Au collisions at $\sqrt{s_{\rm NN}}=$~200~GeV. Approximately 10~M events of Au+Au collisions were generated for each case.


\subsection{Correlators}
The multi-particle correlation methods~\cite{Bilandzic:2010jr,Bilandzic:2013kga,Gajdosova:2017fsc,Jia:2017hbm}, are used in this work. Details on the multi-particle cumulants methods using one- or two-subevents are given in Appendix.~\ref{cum}. In this work the two subevents used with  $\Delta\eta~=~\eta_{1}-\eta_{2}~ > 0.7$  between the subevents $\textit{A}$ and $\textit{B}$ (\textit{i.e.}, $\eta_{A}~ > 0.35$ and $\eta_{B}~ < -0.35$).
The need for high statistical power precludes using the two-subevents method to calculate the five- and six-particle correlations. Consequently, the one-subevent method was used for such calculations.

%
Using multi-particle correlations methods, the Symmetric Correlations and Asymmetric Correlations can be given as:
\begin{widetext}
\begin{itemize}
\item{The two-subevents two-particle $SC(n_1,n_2)$ with ($n_2$ = - $n_1$);}
\end{itemize}
{\small \begin{eqnarray}\label{eq:2-1}
    SC(n_1, -n_1) &=& \langle\langle  \cos(n_{1}\phi^{A}_1 - n_{1}\phi^{B}_2 \rangle\rangle, \\  \nonumber
                  &=& \langle\langle  v_{n_{1}} v_{n_{2}} \cos(n_{1}\psi_{n_{1}} + n_{2}\psi_{n_{2}} )\rangle\rangle, \\  \nonumber
                  &=& \langle\langle  v^{2}_{n_{1}} \rangle\rangle.
\end{eqnarray} }
\begin{itemize}
\item{The two-subevents three-particle $ASC(n_1,n_2,n_3)$ with ($n_3$ = - $n_1$ - $n_2$);}
\end{itemize}
{\small \begin{eqnarray}\label{eq:2-2}
    ASC(n_1, n_2, n_3) &=& \langle\langle  \cos(n_{1}\phi^{A}_1 + n_{2}\phi^{A}_2 + n_{3}\phi^{B}_3 \rangle\rangle, \\  \nonumber
                       &=& \langle\langle  v_{n_{1}} v_{n_{2}} v_{n_{3}} \cos(n_{1}\psi_{n_{1}} + n_{2}\psi_{n_{2}} + n_{3}\psi_{n_{3}} )\rangle\rangle,  \\  \nonumber
                       &=& \langle\langle  v_{n_{1}} v_{n_{2}} v_{n_{1}+n_{2}} \cos(n_{1}\psi_{n_{1}} + n_{2}\psi_{n_{2}} - (n_{1}+ n_{2})\psi_{{n_{1}+n_{2}}} )\rangle\rangle.
\end{eqnarray} }
\begin{itemize}
\item{The two-subevents four-particle $SC(n_1,n_2,n_3,n_4)$ with ($n_3$ = -$n_1$ and $n_4$ = -$n_2$ );}
\end{itemize}
{\small \begin{eqnarray}\label{eq:2-3}
    SC(n_1, n_2, n_3, n_4) &=& \langle\langle  \cos(n_{1}\phi^{A}_1 + n_{2}\phi^{A}_2 + n_{3}\phi^{B}_3 + n_{4}\phi^{B}_4 \rangle\rangle, \\  \nonumber
                       &=& \langle\langle  v_{n_{1}} v_{n_{2}} v_{n_{3}} v_{n_{4}} \cos(n_{1}\psi_{n_{1}} + n_{2}\psi_{n_{2}} + n_{3}\psi_{n_{3}} +  n_{4}\psi_{n_{4}} )\rangle\rangle, \\  \nonumber
                       &=& \langle\langle  v^{2}_{n_{1}}  v^{2}_{n_{2}} \rangle\rangle.
\end{eqnarray}  }
\begin{itemize}
\item{The two-subevents  four-particle $ASC(n_1,n_2,n_3,n_4)$ with ($n_1 + n_2 + n_3 + n_4$ = 0);}
\end{itemize}
{\small \begin{eqnarray}\label{eq:2-4}
   ASC(n_1, n_2, n_3, n_4) &=& \langle\langle  \cos(n_{1}\phi^{A}_1 + n_{2}\phi^{A}_2 + n_{3}\phi^{B}_3 + n_{4}\phi^{B}_4 \rangle\rangle, \\  \nonumber
                       &=& \langle\langle  v_{n_{1}} v_{n_{2}} v_{n_{3}} v_{n_{4}} \cos(n_{1}\psi_{n_{1}} + n_{2}\psi_{n_{2}} + n_{3}\psi_{n_{3}} +  n_{4}\psi_{n_{4}} )\rangle\rangle.
\end{eqnarray} }

\begin{itemize}
\item{The one-subevent five-particle $ASC(n_1,n_2,n_3,n_4,n_5)$ with ($n_1 + n_2 + n_3 + n_4 + n_5$ = 0);}
\end{itemize}
{\small \begin{eqnarray}\label{eq:2-5}
   ASC(n_1, n_2, n_3, n_4, n_5) &=& \langle\langle  \cos(n_{1}\phi_1 + n_{2}\phi_2 + n_{3}\phi_3 + n_{4}\phi_4 + n_{5}\phi_5\rangle\rangle, \\  \nonumber
                       &=& \langle\langle  v_{n_{1}} v_{n_{2}} v_{n_{3}} v_{n_{4}} v_{n_{5}} \cos(n_{1}\psi_{n_{1}} + n_{2}\psi_{n_{2}} + n_{3}\psi_{n_{3}} +  n_{4}\psi_{n_{4}} +  n_{5}\psi_{n_{5}})\rangle\rangle.
\end{eqnarray} }
\begin{itemize}
\item{The one-subevent  six-particle $SC(n_1,n_2,n_3,n_4,n_5,n_6)$ with ($n_4$ = -$n_1$,  $n_5$ = -$n_2$, and $n_6$ = -$n_3$ );}
\end{itemize}
{\small \begin{eqnarray}\label{eq:2-6}
   SC(n_1, n_2, n_3, n_4, n_5, n_6) &=& \langle\langle  \cos(n_{1}\phi_1 + n_{2}\phi_2 + n_{3}\phi_3 + n_{4}\phi_4 + n_{5}\phi_5 + n_{6}\phi_6 \rangle\rangle \\  \nonumber
                                    &=& \langle\langle  v_{n_{1}} v_{n_{2}} v_{n_{3}} v_{n_{4}} v_{n_{5}} v_{n_{6}}  \cos(n_{1}\psi_{n_{1}} + n_{2}\psi_{n_{2}} + n_{3}\psi_{n_{3}} +  n_{4}\psi_{n_{4}} +  n_{5}\psi_{n_{5}} +  n_{6}\psi_{n_{6}} )\rangle\rangle,  \\  \nonumber
                       &=& \langle\langle  v^{2}_{n_{1}}  v^{2}_{n_{2}} v^{3}_{n_{3}} \rangle\rangle.
\end{eqnarray} }

The multi-particle correlations Eqs.~\ref{eq:2-1} and~\ref{eq:2-3} can be used to define the same and mix flow harmonics NSC as:
{\small  \begin{eqnarray}\label{eq:3-4}
\gamma_{n_1,n_1,-n_1,-n_1} &=& \dfrac{SC(n_1,n_1,-n_1,-n_1)}{SC(n_1,-n_1) SC(n_1,-n_1)} -2.
\end{eqnarray}
\begin{eqnarray}\label{eq:3-4-1}
\beta_{n_1,n_2,-n_1,-n_2} &=& \dfrac{SC(n_1,n_2,-n_1,-n_2)}{SC(n_1,-n_1) SC(n_2,-n_2)} -1.
\end{eqnarray} }

In addition, the multi-particle correlations Eqs.~\ref{eq:2-1}--\ref{eq:2-6} can define the Normalized Asymmetric Correlations (flow angular correlations) as:
{\small
\begin{eqnarray}\label{eq:3-1}
\rho_{n_1,n_2,n_3} &=& \dfrac{ASC(n_1,n_2,n_3)}{\sqrt{|SC(n_1,n_2,-n_1,-n_2) SC(n_3, -n_3)|}}, \\  \nonumber
                   &\sim & \langle \cos(n_1 \psi_{n_1} + n_2 \psi_{n_1} + n_3 \psi_{n_3}) \rangle.
\end{eqnarray}
\begin{eqnarray}\label{eq:3-2}
\rho_{n_1,n_2,n_3,n_4} &=& \dfrac{ASC(n_1,n_2,n_3,n_4)}{\sqrt{|SC(n_1,n_2,n_3,-n_1,-n_2,-n_3) SC(n_4, -n_4)|}}, \\  \nonumber
                         &\sim & \langle \cos(n_1 \psi_{n_1} + n_2 \psi_{n_1} + n_3 \psi_{n_3} + n_4 \psi_{n_4}) \rangle.
\end{eqnarray}
\begin{eqnarray}\label{eq:3-3}
\rho_{n_1,n_2,n_3,n_4,n_5} &=& \dfrac{ASC(n_1,n_2,n_3,n_4,n_5)}{\sqrt{|SC(n_1,n_2,n_3,-n_1,-n_2,-n_3) SC(n_4,n_5,-n_4,-n_5)|}}, \\  \nonumber
                         &\sim & \langle \cos(n_1 \psi_{n_1} + n_2 \psi_{n_1} + n_3 \psi_{n_3} + n_4 \psi_{n_4} + n_5 \psi_{n_5}) \rangle.
\end{eqnarray}
}
The flow angular correlations given by the NSC are summarized in Tab~\ref{tab:2}.
\begin{table*}[h!]
\begin{center}
\caption{The summary of the NASC that is presented in this work.\label{tab:2}}
 \begin{tabular}{|c|c|c|c|}
 \hline 
  NASC    &  Event-plane correlations  \\
  \hline
  $\langle \cos(2 \psi_{1} - 2 \psi_{2}) \rangle$                             & $\rho_{1,1,-2}$        \\
  $                                             $                             & $\rho_{1,1,1,-1,-2}$    \\
  \hline
  $\langle \cos(6 \psi_{2} - 6 \psi_{3}) \rangle$                             & $\rho_{2,2,2,-3,-3}$    \\
  \hline
  $                                             $                             & $\rho_{2,2,-4}$          \\
  $\langle \cos(4 \psi_{2} - 4 \psi_{4}) \rangle$                             & $\rho_{2,2,2,-2,-4}$     \\
  $                                             $                             & $\rho_{2,2,3,-3,-4}$      \\
  \hline
  $\langle \cos(6 \psi_{2} - 6 \psi_{6}) \rangle$                             & $\rho_{2,2,2,-6}$      \\
  \hline
  $\langle \cos(1 \psi_{1} + 2 \psi_{2} - 3 \psi_{3}) \rangle$                & $\rho_{1,2,-3}$        \\
  \hline
  $\langle \cos(1 \psi_{1} + 3 \psi_{2} - 4 \psi_{3} \rangle$                 & $\rho_{1,2,-4}$        \\
  \hline
  $                                                          $                & $\rho_{2,3,-5}$         \\
  $\langle \cos(2 \psi_{2} + 3 \psi_{3} - 5 \psi_{5}) \rangle$                & $\rho_{2,3,2,-2,-5}$     \\
  $                                                          $                & $\rho_{2,3,3,-3,-5}$      \\
  \hline
  $\langle \cos(6 \psi_{3} - 2 \psi_{2} - 4 \psi_{4}) \rangle$                & $\rho_{3,3,-2,-4}$         \\
  $\langle \cos(2 \psi_{2} + 6 \psi_{3} - 8 \psi_{4} ) \rangle$               & $\rho_{2,3,3,-4,-4}$        \\
  \hline
  $\langle \cos(2 \psi_{2} + 5 \psi_{5} - 3 \psi_{3} - 4 \psi_{4}) \rangle$   & $\rho_{2,5,-3,-4}$          \\
  $\langle \cos(4 \psi_{2} + 4 \psi_{4} - 3 \psi_{3} - 5 \psi_{5} ) \rangle$  & $\rho_{2,2,4,-3,-5}$         \\
  \hline
\end{tabular}
\end{center}
\end{table*}
\end{widetext}
\section{Results and discussion}\label{sec:3}
Non-flow effects can impact the reliability of the extracted flow correlations. Therefore, it is informative to consider a figure of merit for these contributions to the multi-particle correlations. The HIJING model results in Appendix~\ref{Hijing} suggest that the two-, three- and four-particle correlations (two-subevents) and the five- and six-particle correlations (one-subevent) have little (if any)  non-flow contributions.
{\color{black}One caveat to the use of the HIJING model is that it does not consider the medium’s response to the non-flow effects (see Ref~\cite{Pablos:2022piv}).}
This section will discuss the final state effects on the symmetric (asymmetric) correlations and the flow magnitudes and angular correlations.
\subsection{Symmetric correlations}
\begin{figure}[!h] 
\includegraphics[width=1.0  \linewidth, angle=-0,keepaspectratio=true,clip=true]{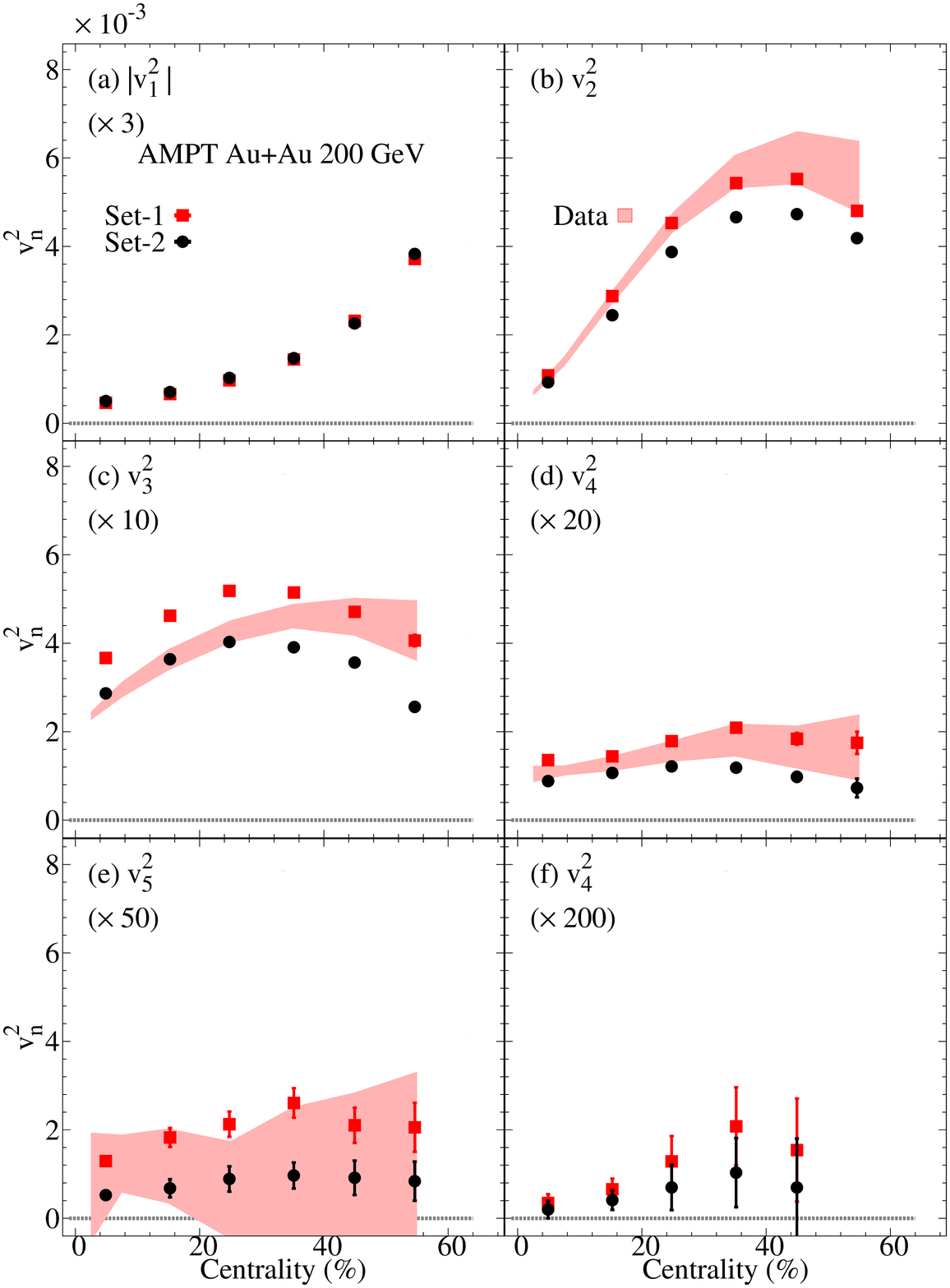}
\vskip -0.4cm
\caption{ 
The centrality dependence of the two-particle flow harmonics $v^{2}_{n} = SC$($n$,$-n$) using the two-subevents method for Au+Au at $\sqrt{\textit{s}_{NN}}~=$ 200~GeV from the AMPT model Set-1 and Set-2. The solid curves represent the experimental data~\cite{Adamczyk:2016exq,Adamczyk:2017hdl}.
}\label{fig:c1}
\vskip -0.3cm
\end{figure}
\begin{figure}[!h] 
\includegraphics[width=0.80 \linewidth, angle=-0,keepaspectratio=true,clip=true]{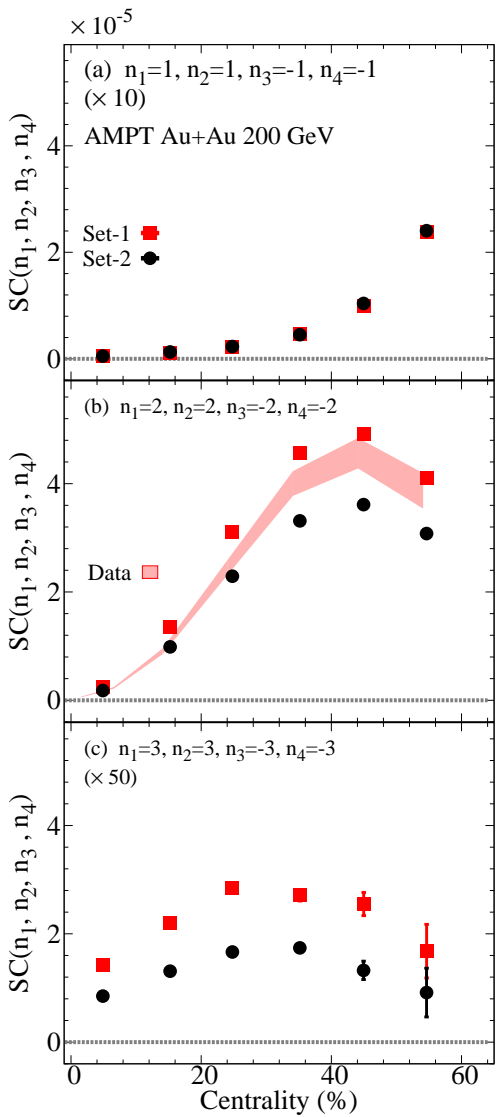}
\vskip -0.4cm
\caption{
The centrality dependence of the four-particle symmetric correlations $SC$($1$,$1$,$-1$,$-1$) panel (a),  $SC$($2$,$2$,$-2$,$-2$) panel (b) and  $SC$($3$,$3$,$-3$,$-3$) panel (c), using the two-subevents method for Au+Au at $\sqrt{\textit{s}_{NN}}~=$ 200~GeV from the AMPT model Set-1 and Set-2. The solid curve represents the experimental data constructed from Ref.~\cite{STAR:2020gcl}.
}\label{fig:c2}
\vskip -0.3cm
\end{figure}
\begin{figure}[!h] 
\includegraphics[width=0.80 \linewidth, angle=-0,keepaspectratio=true,clip=true]{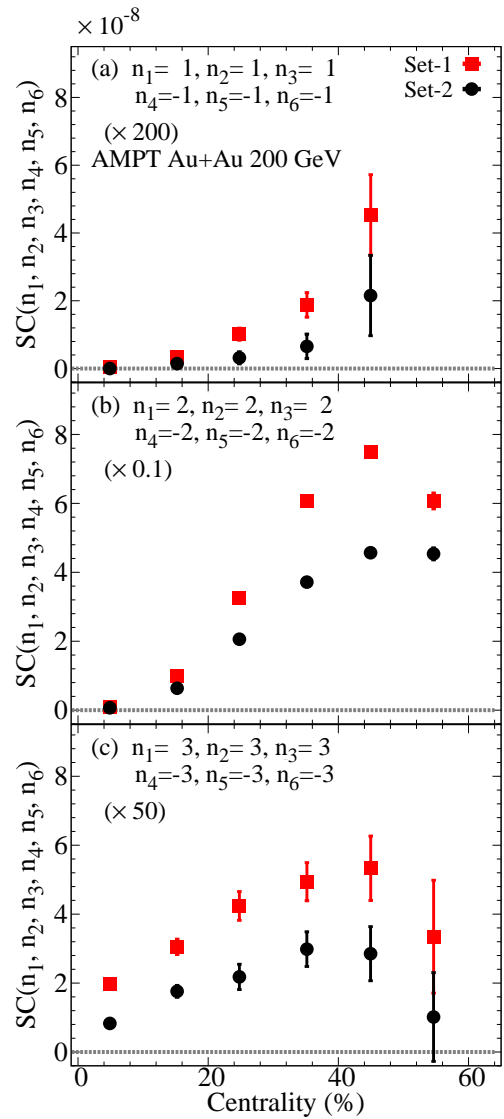}
\vskip -0.4cm
\caption{
The centrality dependence of the six-particle symmetric correlations $SC$($1$,$1$,$1$,$-1$,$-1$,$-2$) panel (a),  $SC$($2$,$2$,$2$,$-2$,$-2$,$-2$) panel (b) and  $SC$($3$,$3$,$3$,$-3$,$-3$,$-3$) panel (c), using the one-subevent method for Au+Au at $\sqrt{\textit{s}_{NN}}~=$ 200~GeV from the AMPT model Set-1 and Set-2.
}\label{fig:c3}
\vskip -0.3cm
\end{figure}

The same harmonic two-, four-, and six-particle symmetric correlations' dependence on the centrality and final state effects are shown in Figs.~\ref{fig:c1},~\ref{fig:c2}, and~\ref{fig:c3}, respectively, for Au+Au collisions at 200~GeV from the AMPT model.
The same harmonic four-and six-particle correlations are expected to contain the contributions of flow and flow fluctuations (i.e., initial state density fluctuations and hydrodynamic evolution fluctuations).
My results show that the same harmonic correlations (n$>$1) decrease with the $\sigma_{pp}$. This characteristic dependence indicates the same harmonic correlations sensitivity to the final state effects given by the AMPT model. 
Note that the same harmonic correlations, with n$=$1, are expected to have a significant global momentum conservation (GMC) effect~\cite{STAR:2018gji,STAR:2019zaf}. The GMC effect on such correlations will be considered in future work. 
The calculations indicate similar values and trends to the experimental data taken/ constructed from Refs.~\cite{Adamczyk:2016exq,Adamczyk:2017hdl,STAR:2020gcl,Magdy:2020bij} (solid curve). Comparisons between the  AMPT calculations and the available experimental measurements of $v^2_n$ (Fig.~\ref{fig:c1}) and $SC$($2$,$2$,$-2$,$-2$) (Fig.~\ref{fig:c2}) reveal the ability of the same harmonic correlations to constrain the AMPT final state effects.

\begin{figure}[!h] 
\includegraphics[width=0.80 \linewidth, angle=-0,keepaspectratio=true,clip=true]{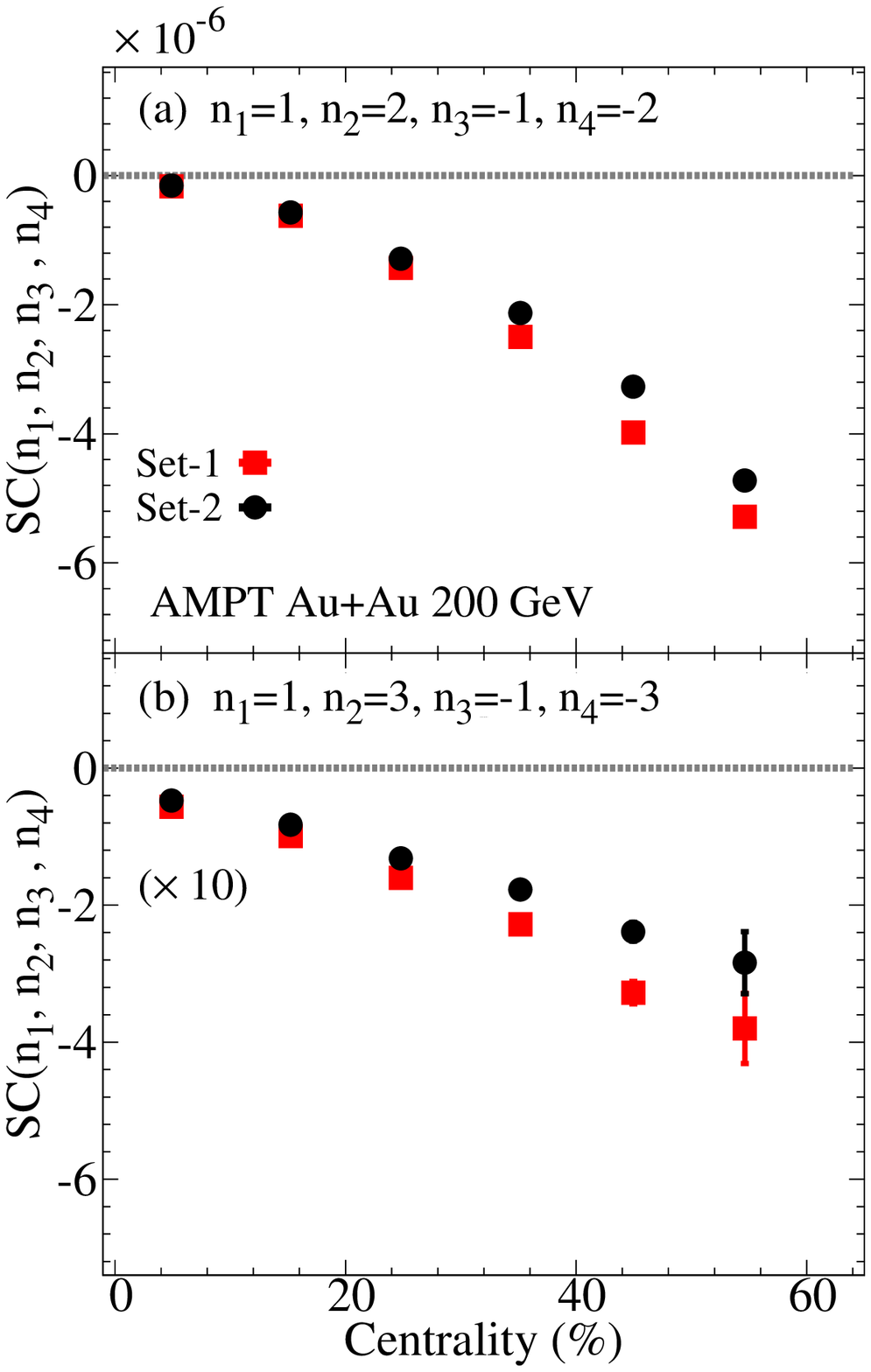}
\vskip -0.4cm
\caption{
The centrality dependence of the four-particle symmetric correlations $SC$($1$,$2$,$-1$,$-2$) panel (a) and $SC$($1$,$3$,$-1$,$-3$) panel (b), using the two-subevents method for Au+Au at $\sqrt{\textit{s}_{NN}}~=$ 200~GeV from the AMPT model Set-1 and Set-2.
}\label{fig:c4}
\vskip -0.3cm
\end{figure}
\begin{figure}[!h] 
\includegraphics[width=0.80 \linewidth, angle=-0,keepaspectratio=true,clip=true]{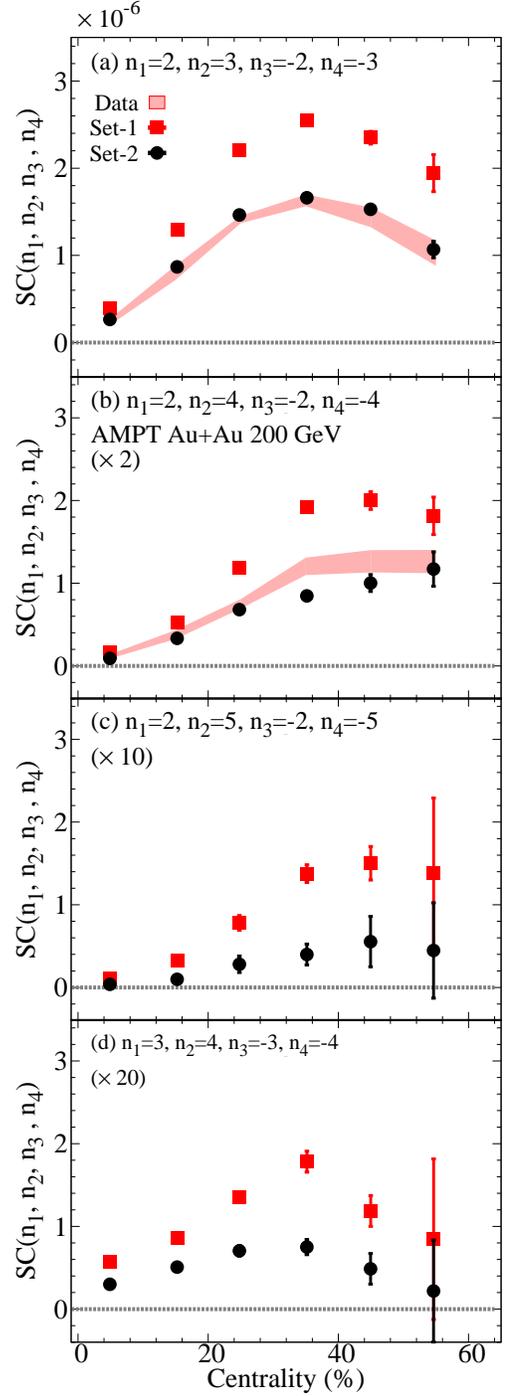}
\vskip -0.4cm
\caption{
The centrality dependence of the four-particle symmetric correlations $SC$($2$,$3$,$-2$,$-3$) panel (a),  $SC$($2$,$4$,$-2$,$-4$) panel (b), $SC$($2$,$4$,$-2$,$-4$) panel (c), and $SC$($3$,$4$,$-3$,$-4$) panel (d), using the two-subevents method for Au+Au at $\sqrt{\textit{s}_{NN}}~=$ 200~GeV from the AMPT model Set-1 and Set-2. The solid curves represent the experimental data constructed from Ref~\cite{STAR:2018fpo}.
}\label{fig:c5}
\vskip -0.3cm
\end{figure}
\begin{figure}[!h] 
\includegraphics[width=0.80 \linewidth, angle=-0,keepaspectratio=true,clip=true]{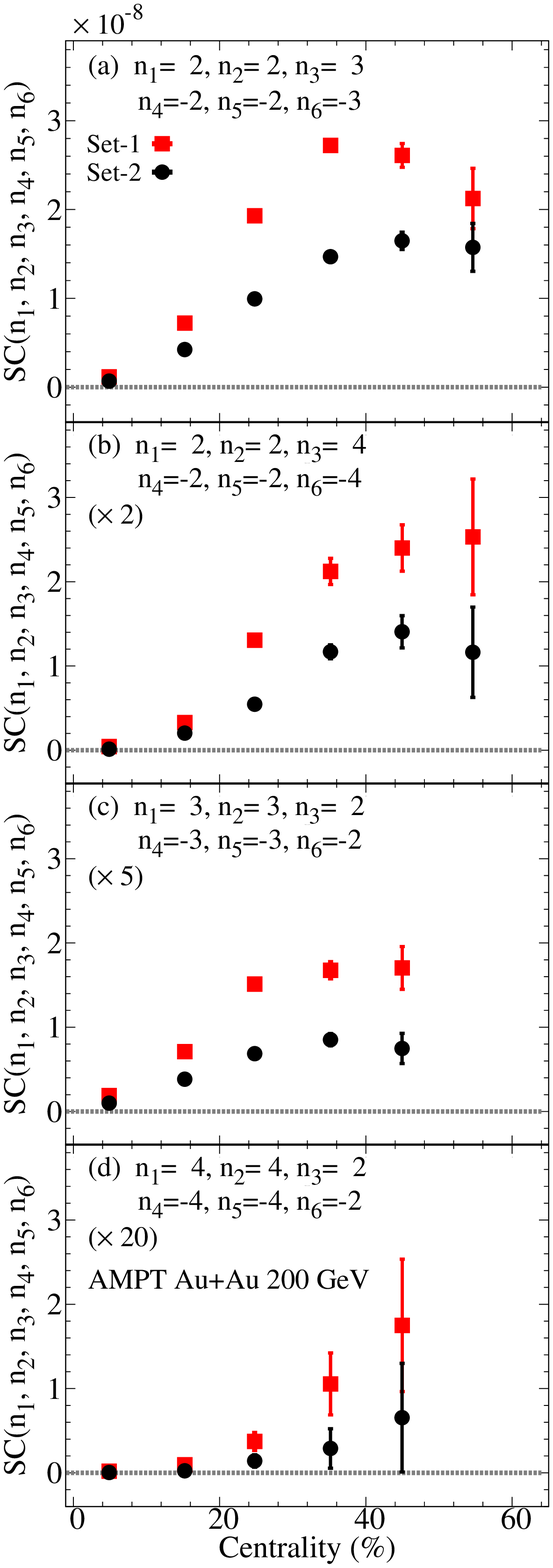}
\vskip -0.4cm
\caption{
The centrality dependence of the six-particle symmetric correlations $SC$($2$,$2$,$3$,$-2$,$-2$,$-3$) panel (a),  $SC$($2$,$2$,$4$,$-2$,$-2$,$-4$) panel (b),  $SC$($3$,$3$,$2$,$-3$,$-3$,$-2$) panel (c), and $SC$($4$,$4$,$2$,$-4$,$-4$,$-2$) panel (d),  using the one-subevent method for Au+Au at $\sqrt{\textit{s}_{NN}}~=$ 200~GeV from the AMPT model Set-1 and Set-2.
}\label{fig:c6}
\vskip -0.3cm
\end{figure}
\begin{figure}[!h] 
\includegraphics[width=0.80 \linewidth, angle=-0,keepaspectratio=true,clip=true]{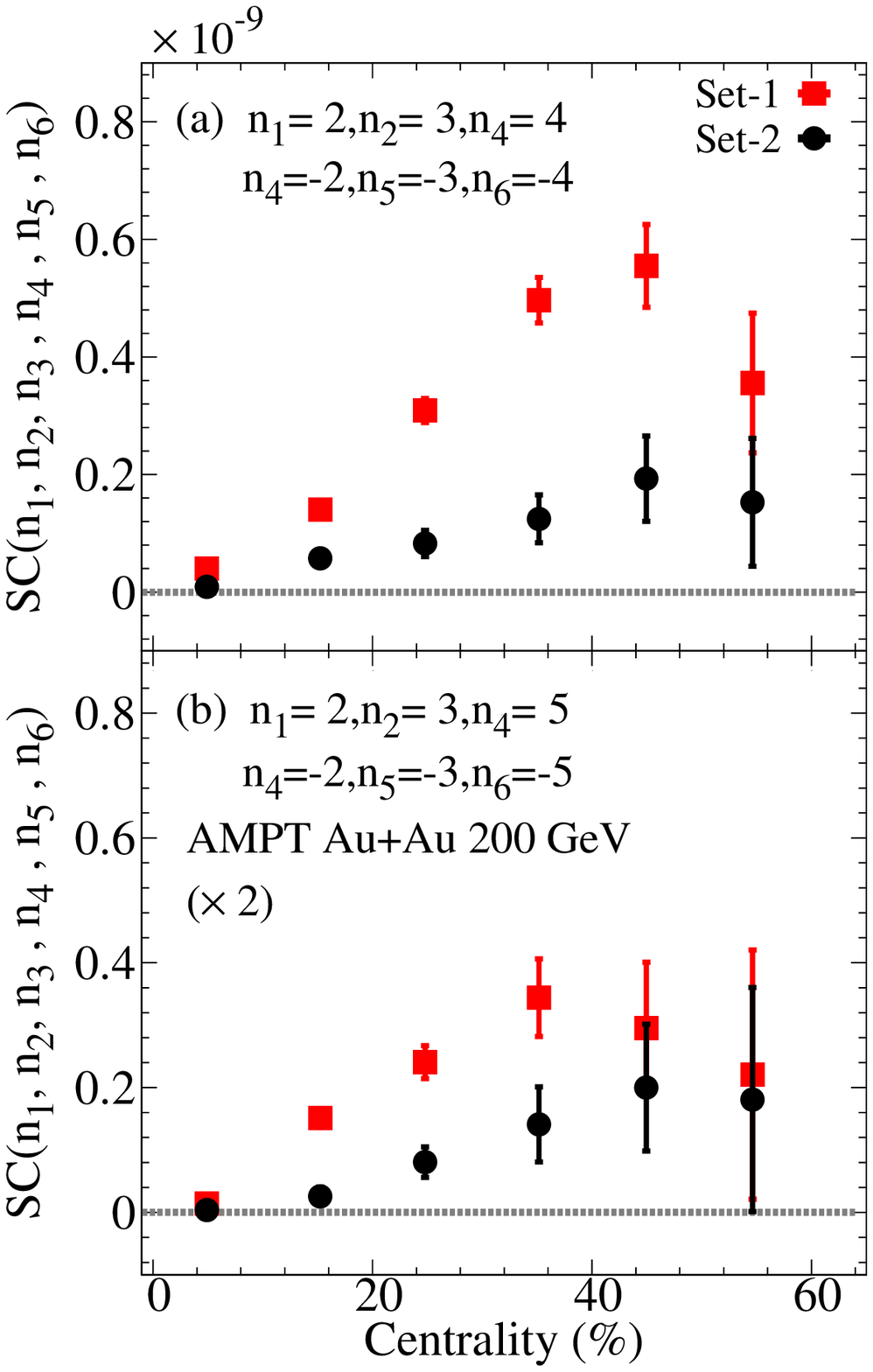}
\vskip -0.4cm
\caption{
The centrality dependence of the six-particle symmetric correlations $SC$($2$,$3$,$4$,$-2$,$-3$,$-4$) panel (a) and  $SC$($2$,$3$,$5$,$-2$,$-3$,$-5$) panel (b), using the one-subevent method for Au+Au at $\sqrt{\textit{s}_{NN}}~=$ 200~GeV from the AMPT model Set-1 and Set-2.
}\label{fig:c7}
\vskip -0.3cm
\end{figure}

The mixed harmonics four- and six-particle symmetric correlations are given in Figs.~\ref{fig:c4},~\ref{fig:c5},~\ref{fig:c6}, and~\ref{fig:c7}, as a function of centrality for Au+Au collisions at 200~GeV from the AMPT model.
The mixed harmonics four- and six-particle correlations are expected to give the flow harmonics correlations induced by initial and final state effects. These correlations will be given using the ratios defined in Eq.~\ref{eq:3-4-1}.
The AMPT calculations indicate that mixed harmonics correlations decrease with the $\sigma_{pp}$, representing the mixed harmonics correlations' sensitivity to the final state effects given by the AMPT model. 
The four-particle symmetric correlations with $n_1 = 1$, Fig.~\ref{fig:c4}, are expected to have a contributions form GMC~\cite{STAR:2018gji,STAR:2019zaf}.
In addition, my calculations in Fig.~\ref{fig:c5} show similar values and behaviors to the experimental data constructed from Ref.~\cite{STAR:2018fpo} (solid curve). 
Therefore, this work suggests that more detailed data-model comparisons can better constrain the final state effects.


\subsection{Asymmetric correlations}
\begin{figure}[!h] 
\includegraphics[width=0.80 \linewidth, angle=-0,keepaspectratio=true,clip=true]{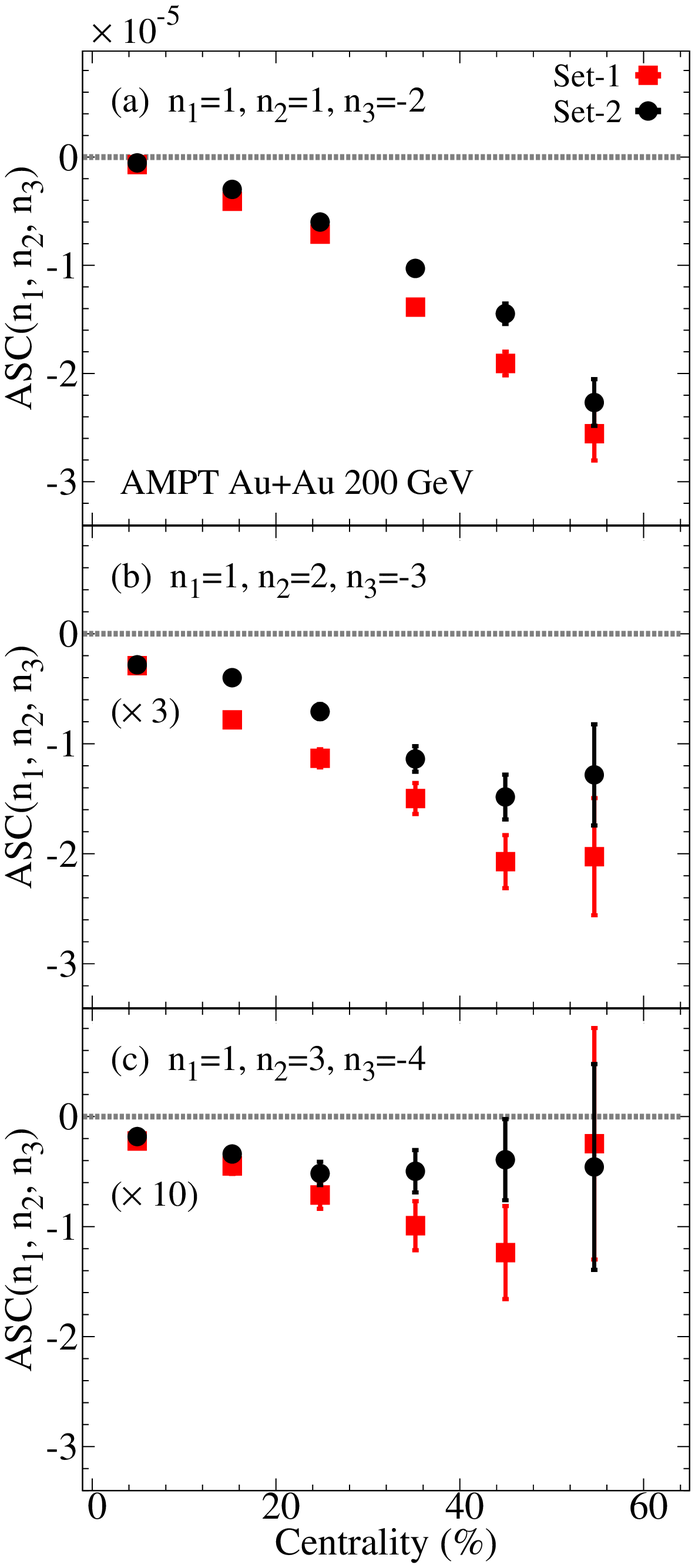}
\vskip -0.4cm
\caption{
The centrality dependence of the three-particle asymmetric correlations $ASC$($1$,$1$,$-2$) panel (a),  $ASC$($1$,$2$,$-3$) panel (b) and  $ASC$($1$,$3$,$-4$) panel (c),  using two-subevents method for Au+Au at $\sqrt{\textit{s}_{NN}}~=$ 200~GeV from the AMPT model Set-1 and Set-2.
}\label{fig:c8}
\vskip -0.3cm
\end{figure}
\begin{figure}[!h] 
\includegraphics[width=0.80 \linewidth, angle=-0,keepaspectratio=true,clip=true]{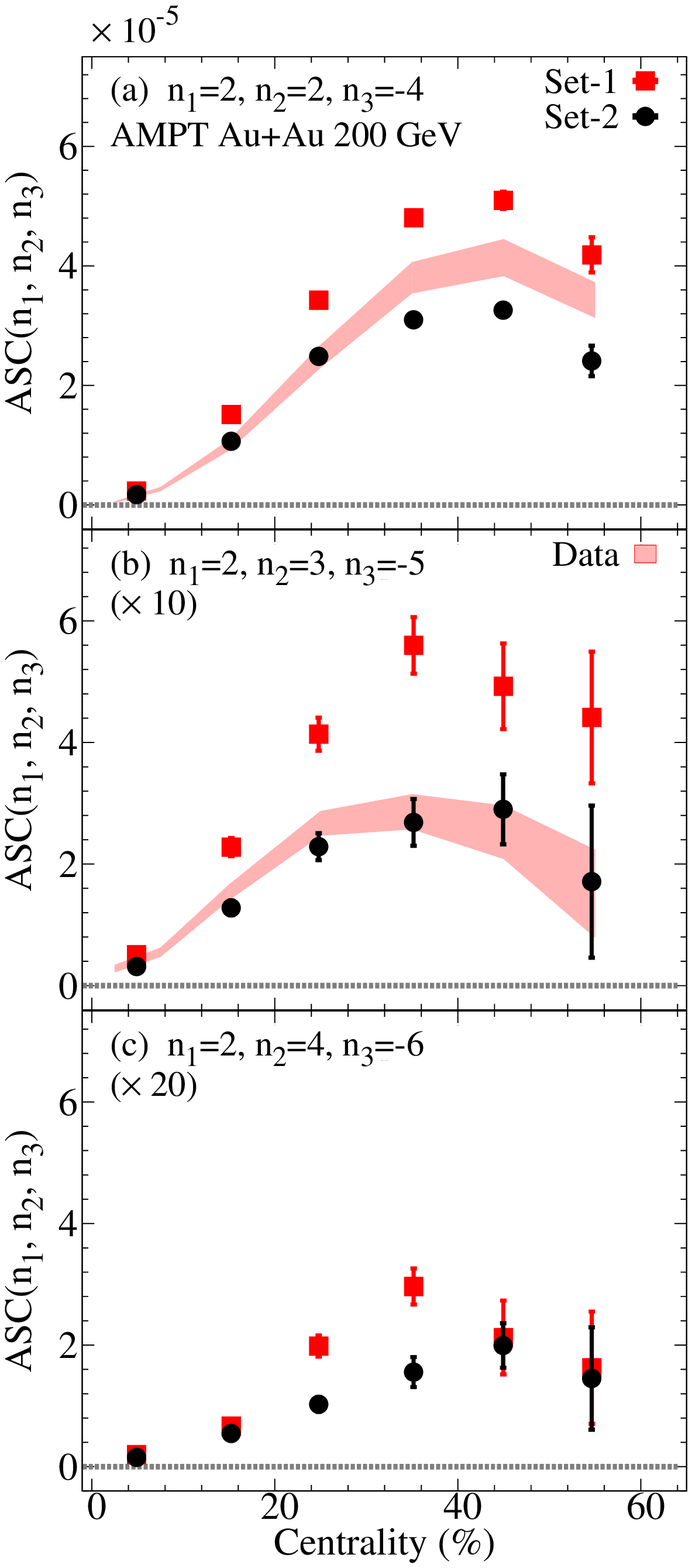}
\vskip -0.4cm
\caption{
The centrality dependence of the three-particle asymmetric correlations $ASC$($2$,$2$,$-4$) panel (a),  $ASC$($2$,$3$,$-5$) panel (b) and  $ASC$($2$,$4$,$-5$) panel (c), using two-subevents method for Au+Au at $\sqrt{\textit{s}_{NN}}~=$ 200~GeV from the AMPT model Set-1 and Set-2. The solid curves represent the experimental data~\cite{STAR:2020gcl}.
}\label{fig:c9}
\vskip -0.3cm
\end{figure}
\begin{figure}[!h] 
\includegraphics[width=0.80 \linewidth, angle=-0,keepaspectratio=true,clip=true]{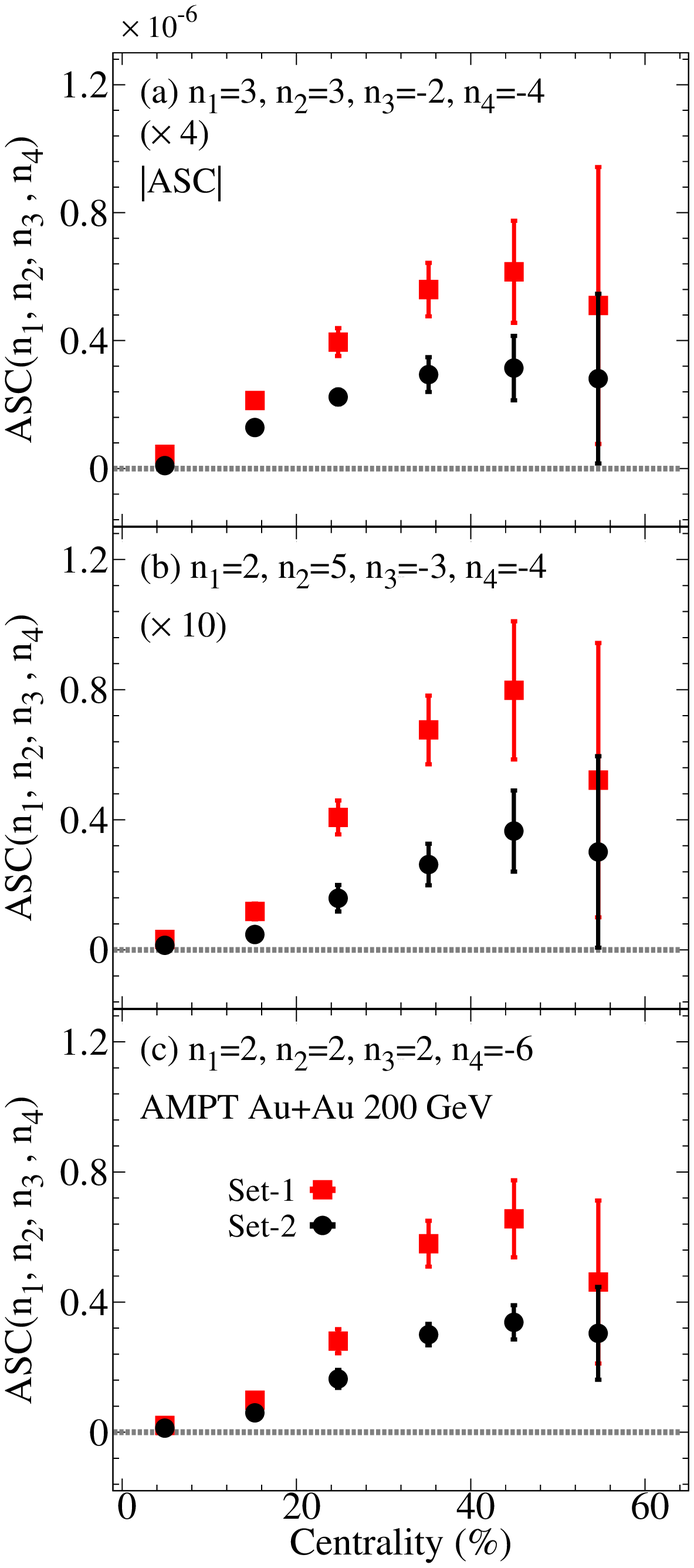}
\vskip -0.4cm
\caption{
The centrality dependence of the four-particle asymmetric correlations $ASC$($3$,$3$,$-2$,$-4$) panel (a),  $ASC$($2$,$5$,$-3$,$-4$) panel (b) and  $ASC$($2$,$2$,$2$,$-6$) panel (c), using two-subevents method for Au+Au at $\sqrt{\textit{s}_{NN}}~=$ 200~GeV from the AMPT model Set-1 and Set-2.
}\label{fig:c10}
\vskip -0.3cm
\end{figure}
\begin{figure}[!h] 
\includegraphics[width=0.80 \linewidth, angle=-0,keepaspectratio=true,clip=true]{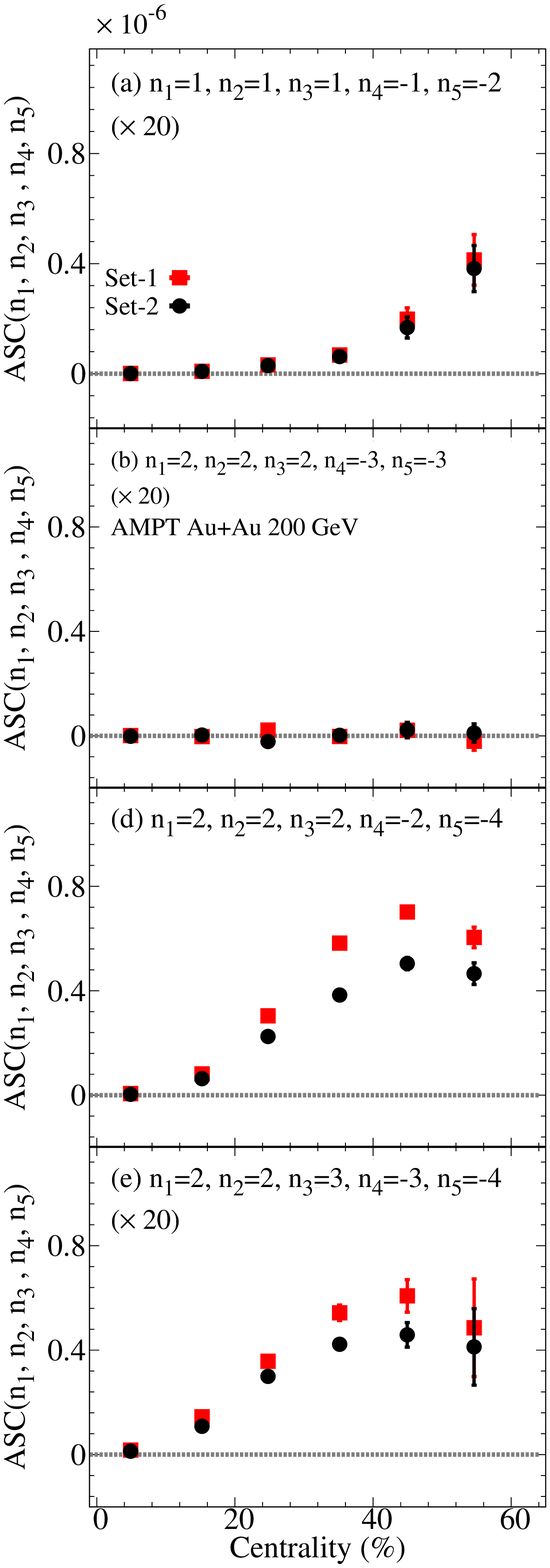}
\vskip -0.4cm
\caption{
The centrality dependence of the five-particle asymmetric correlations $ASC$($1$,$1$,$1$,$-1$,$-2$) panel (a),  $ASC$($2$,$2$,$2$,$-3$,$-3$) panel (b), $ASC$($2$,$2$,$2$,$-2$,$-4$) panel (c), and $ASC$($2$,$2$,$3$,$-3$,$-4$) panel (d), using one-subevent method for Au+Au at $\sqrt{\textit{s}_{NN}}~=$ 200~GeV from the AMPT model Set-1 and Set-2.
}\label{fig:c11}
\vskip -0.3cm
\end{figure}
\begin{figure}[!h] 
\includegraphics[width=0.80 \linewidth, angle=-0,keepaspectratio=true,clip=true]{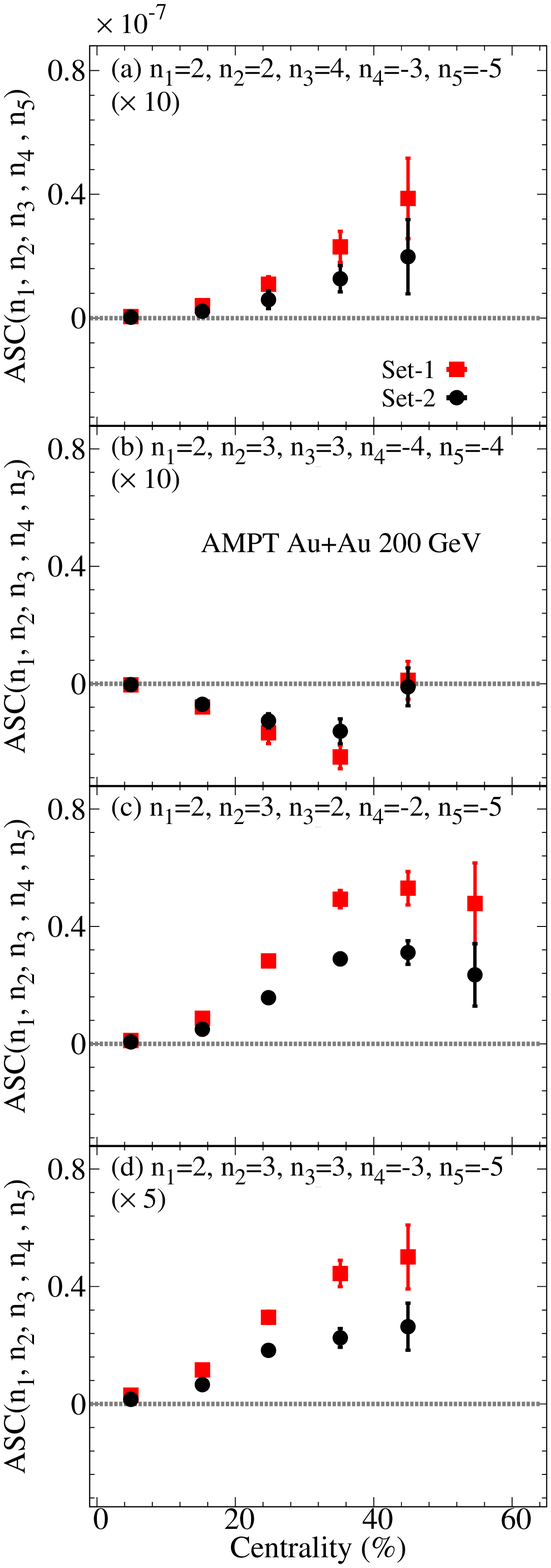}
\vskip -0.4cm
\caption{
The centrality dependence of the five-particle asymmetric correlations $ASC$($2$,$2$,$4$,$-3$,$-5$) panel (a),  $ASC$($2$,$3$,$3$,$-4$,$-4$) panel (b), $ASC$($2$,$3$,$2$,$-2$,$-5$) panel (c), and $ASC$($2$,$3$,$3$,$-3$,$-5$) panel (d), using one-subevent method for Au+Au at $\sqrt{\textit{s}_{NN}}~=$ 200~GeV from the AMPT model Set-1 and Set-2.
}\label{fig:c12}
\vskip -0.3cm
\end{figure}
The three-, four-, and six-particle asymmetric correlations for the AMPT model Au+Au collisions at 200~GeV are shown in Figs.~\ref{fig:c8}--~\ref{fig:c12}. 
The asymmetric correlations can give the flow angular correlations induced by the initial state and summarized in Tab.~\ref{tab:2}.
The AMPT asymmetric correlations given in Figs.~\ref{fig:c8}--~\ref{fig:c12} show that the correlations strength decrease with the $\sigma_{pp}$, which give their sensitivity to the AMPT model final state effects. 
The three- and five-particle asymmetric correlations with $n_i = 1$, Figs.~\ref{fig:c8} and ~\ref{fig:c11}, are expected to be impacted by the GMC effect~\cite{STAR:2018gji,STAR:2019zaf}.
The $ASC$($2$,$2$,$2$,$-3$,$-3$) given in Fig.~\ref{fig:c12} panel (b) show values consistent with zero, reflecting the weak nature of the correlations between $\psi_{2}$ and $\psi_3$.
In addition, the AMPT calculations in Fig.~\ref{fig:c9} show similar values and behaviors to the experimental data reported in Ref.~\cite{STAR:2020gcl} (solid curve). 
Therefore, more detailed data-model comparisons can help better understand the initial state angular correlations. 

\subsection{Normalized symmetric correlations}
The normalized symmetric correlations (i.e., same harmonic flow fluctuations and mixed harmonics flow correlations) are given in Eqs.~\ref{eq:3-4} and ~\ref{eq:3-4-1}. The ratios $\gamma_{n,n,-n,-n}$ serve as a metric for the n$^{th}$-order flow harmonic fluctuations. On the other hand, the $\beta_{n,m,-n,-m}$ ratios measure the strength of mixed harmonics flow correlations.
\begin{figure}[!h] 
\includegraphics[width=0.80 \linewidth, angle=-0,keepaspectratio=true,clip=true]{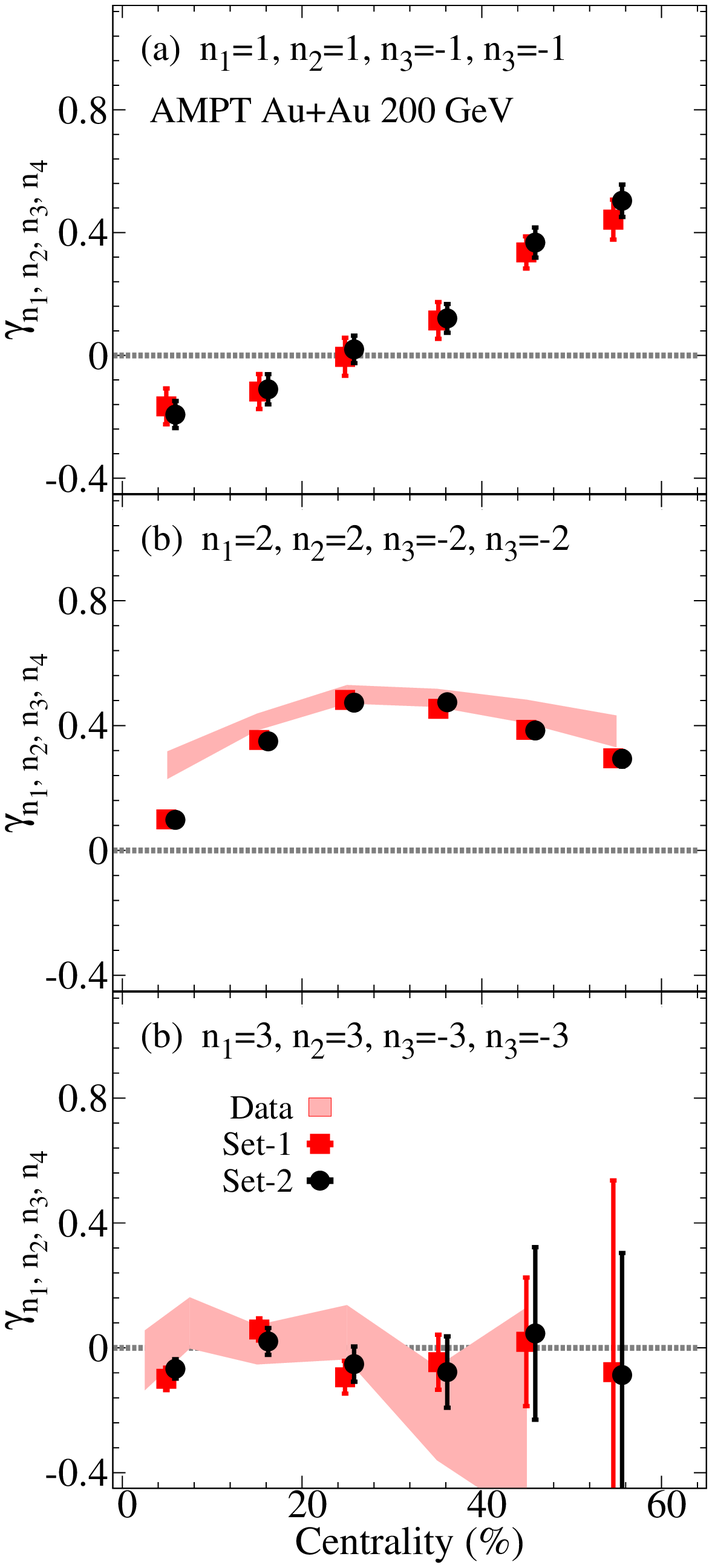}
\vskip -0.4cm
\caption{
The centrality dependence of the same harmonic normalized symmetric correlations $\gamma_{1,1,-1,-1}$ panel (a), $\gamma_{2,2,-2,-2}$ panel (b) and $\gamma_{3,3,-3,-3}$ panel (c), for Au+Au at $\sqrt{\textit{s}_{NN}}~=$ 200~GeV from the AMPT model Set-1 and Set-2. The solid curves represent the experimental data constructed from Refs~\cite{STAR:2019zaf,STAR:2013qio}.
}\label{fig:r1}
\vskip -0.3cm
\end{figure}
\begin{figure}[!h] 
\includegraphics[width=1.0 \linewidth, angle=-0,keepaspectratio=true,clip=true]{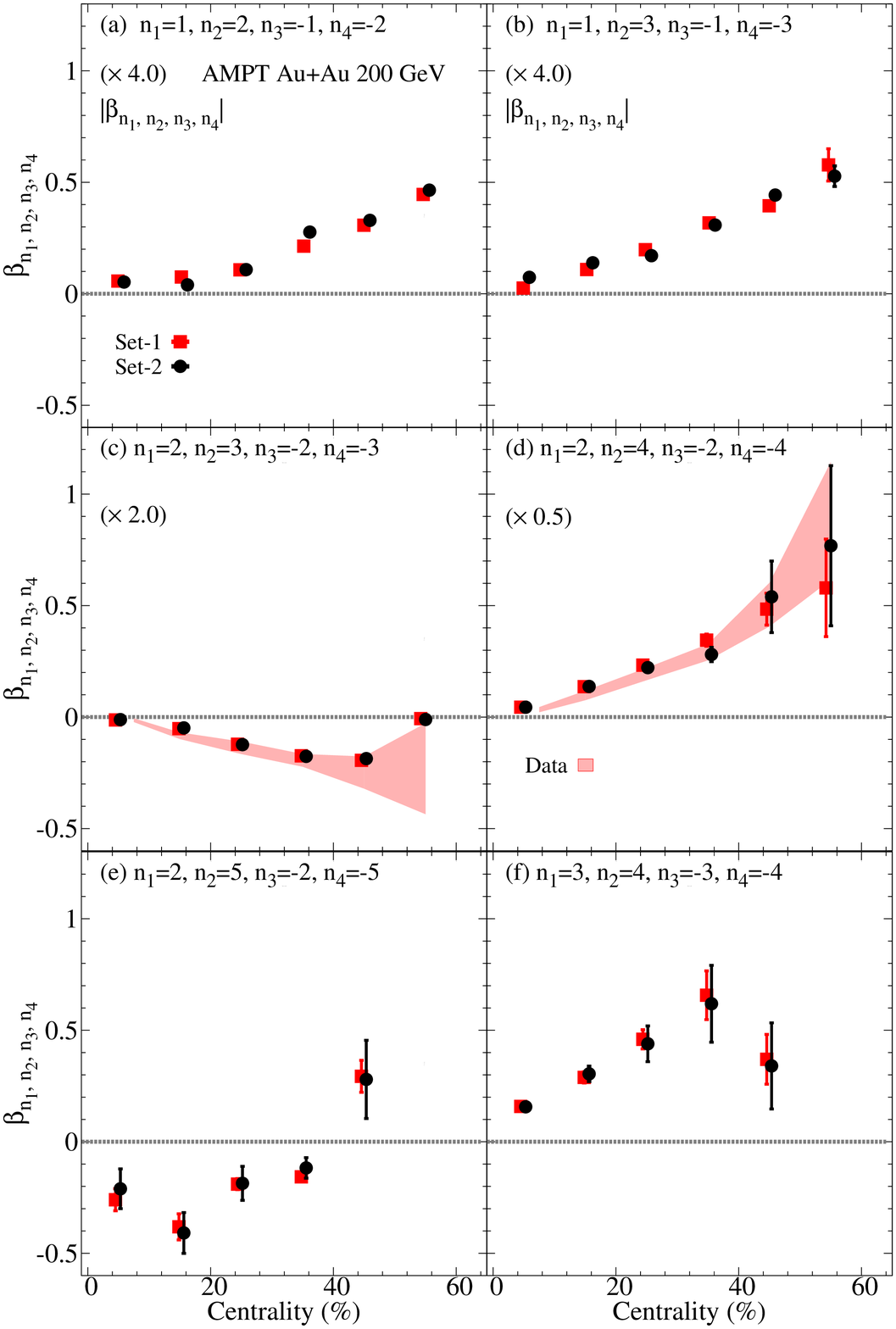}
\vskip -0.4cm
\caption{
The centrality dependence of the normalized three-particle asymmetric correlations $\beta_{1,2,-1,-2}$ panel (a), $\beta_{1,3,-1,-3}$ panel (b), $\beta_{2,3,-2,-3}$ panel (c), $\beta_{2,4,-2,-4}$ panel (d), $\beta_{2,5,-2,-5}$ panel (e), and $\beta_{3,4,-3,-4}$ panel (f) for Au+Au at $\sqrt{\textit{s}_{NN}}~=$ 200~GeV from the AMPT model Set-1 and Set-2. The solid curves represent the experimental data~\cite{STAR:2018fpo}.
}\label{fig:r2}
\vskip -0.3cm
\end{figure}

Figure~\ref{fig:r1} show the normalized symmetric correlations $\gamma_{1,1,-1,-1}$ panel (a), $\gamma_{2,2,-2,-2}$ panel (b) and $\gamma_{3,3,-3,-3}$ panel (c) dependence on the centrality and final state effects given by the AMPT model Set-1 and Set-2. 
The $\gamma_{1,1,-1,-1}$ shows the rapidity-even dipolar flow fluctuations nature in the AMPT model, and it assumes similar GMC effects on the two- and four-particle correlations.
The negative values suggest that flow fluctuations dominate central $\gamma_{1,1,-1,-1}$; however better understanding of the GMC effects is required before further interpretation.
The ratio $\gamma_{2,2,-2,-2}$ gives the centrality dependence of the elliptic flow fluctuations in the AMPT model; it shows the expected decrease in the magnitude of the fluctuations from central to peripheral collisions. The data-model comparison of $\gamma_{2,2,-2,-2}$ indicates that the calculated elliptic flow fluctuations agree with the experimental data~\cite{STAR:2019zaf}; however, it over-predicts the measurements in the central collision. 
The ratio $\gamma_{3,3,-3,-3}$ gives the triangular flow fluctuations. The $\gamma_{3,3,-3,-3}$ from the AMPT model are consistent with zero at all presented centrality selections. Such an observation is consistent with the experimental measurements~\cite{STAR:2013qio}. Therefore, experimental measurements and the AMPT calculations suggest that Gaussian flow fluctuations dominate triangular flow.

The mixed harmonics flow correlations  $\beta_{1,2,-1,-2}$, $\beta_{1,3,-1,-3}$, $\beta_{2,3,-2,-3}$, $\beta_{2,4,-2,-4}$, $\beta_{2,5,-2,-5}$, and $\beta_{3,4,-3,-4}$ from the AMPT model are shown in Fig.~\ref{fig:r2} panels (a)--(f), respectively. 
The $\beta_{1,2,-1,-2}$ and $\beta_{1,3,-1,-3}$ give the rapidity-even dipolar flow $v_1$ correlations with the elliptic and triangular flow $v_2$ and $v_3$, respectively. The calculations show anticorrelations between $v_1$ and $v_n$ ($n$=2,3). 
 On the other hand, $\beta_{2,3,-2,-3}$, $\beta_{2,4,-2,-4}$ and $\beta_{2,5,-2,-5}$ give the correlations between $v_2$ and $v_n$ (n$=$3,4, and 5); however, $\beta_{3,4,-3,-4}$ represents the correlations between $v_3$ and $v_4$.
My results indicated anticorrelations between $v_2$ and $v_3$ and $v_2$ and $v_5$. In contrast, positive correlations are observed between $v_2$ and $v_4$ and between $v_3$ and $v_4$. 
The data-model comparisons of $\beta_{2,3,-2,-3}$ and $\beta_{2,4,-2,-4}$ display that the calculated flow harmonics correlations agree with the experimental data~\cite{STAR:2018fpo}.

The presented $\gamma_{n,n,-n,-n}$ and $\beta_{n,m,-n,-m}$ in Fig.~\ref{fig:r1} and~\ref{fig:r2} show no sensitivity to the $\sigma_{pp}$ (i.e., final state effects). These observations indicate that $\gamma_{n,n,-n,-n}$ and $\beta_{n,m,-n,-m}$  can be used to constrain the initial state fluctuations and correlations, respectively.

\subsection{Normalized asymmetric correlations}
The normalized asymmetric correlations give the flow angular correlations and are provided in Tab.~\ref{tab:2}.
The ratios $\rho_{X}$ serve as a metric for the correlations between flow symmetry planes $\psi_{1}$--$\psi_{6}$.
\begin{figure}[!h] 
\includegraphics[width=0.80 \linewidth, angle=-0,keepaspectratio=true,clip=true]{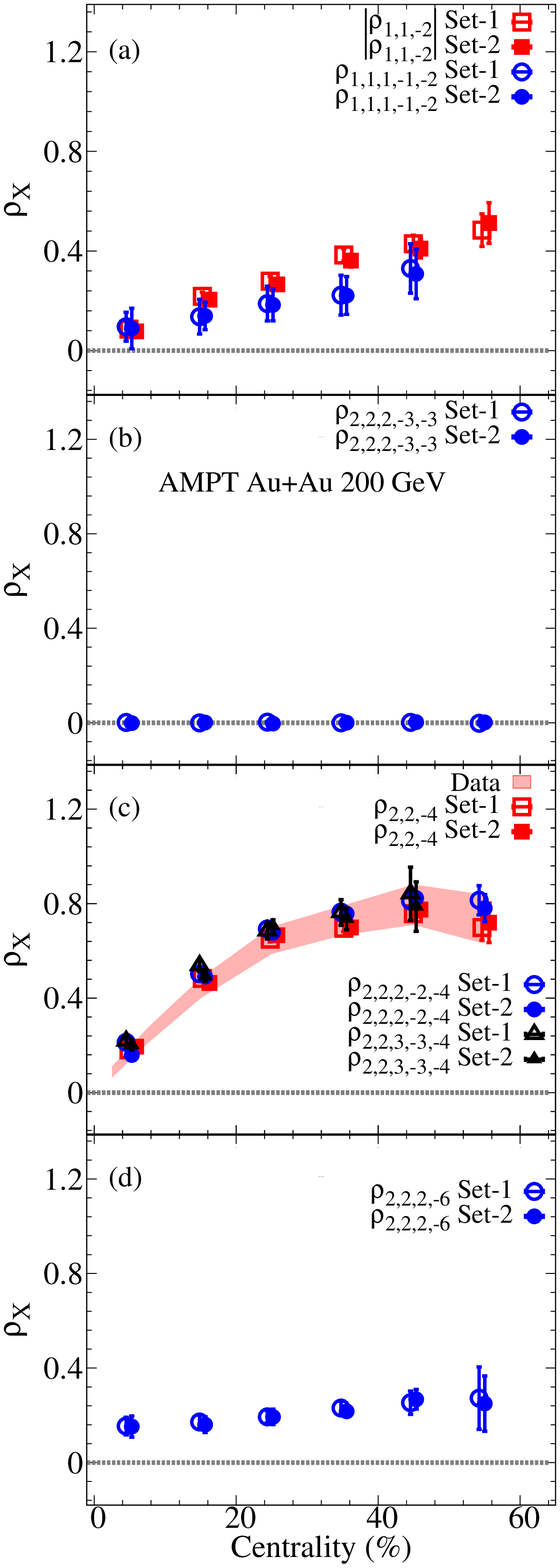}
\vskip -0.4cm
\caption{
The centrality dependence of the normalized asymmetric correlations $\rho_{1,1,-2}$ and $\rho_{1,1,1,-1,2}$ panel (a), $\rho_{2,2,2,-3,-3}$ panel (b), $\rho_{2,2,-4}$, $\rho_{2,2,2,-2,-4}$, and $\rho_{2,2,3,-3,-4}$ panel (c), and $\rho_{2,4,-6}$ panel (d),  for Au+Au at $\sqrt{\textit{s}_{NN}}~=$ 200~GeV from the AMPT model Set-1 and Set-2. The solid curves represent the experimental data~\cite{STAR:2020gcl}.
}\label{fig:r3}
\vskip -0.3cm
\end{figure}
\begin{figure}[!h] 
\includegraphics[width=0.80 \linewidth, angle=-0,keepaspectratio=true,clip=true]{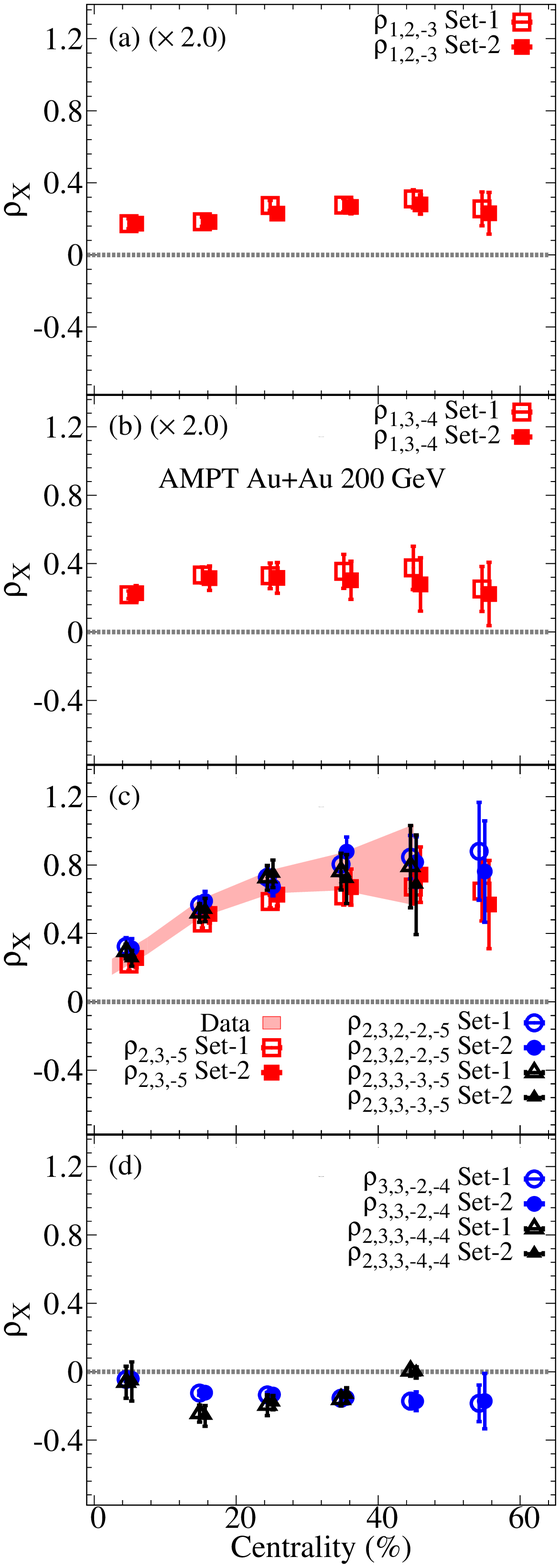}
\vskip -0.4cm
\caption{
The centrality dependence of the normalized asymmetric correlations $\rho_{1,2,-3}$ panel (a), $\rho_{1,3,-4}$ panel (b), $\rho_{2,3,-5}$, $\rho_{2,3,2,-2,-5}$, and $\rho_{2,3,3,-3,-5}$ panel (c), and $\rho_{3,3,-2,-4}$ and $\rho_{2,3,3,-4,-4}$ panel (d),  for Au+Au at $\sqrt{\textit{s}_{NN}}~=$ 200~GeV from the AMPT model Set-1 and Set-2. The solid curves represent the experimental data~\cite{STAR:2020gcl}.
}\label{fig:r4}
\vskip -0.3cm
\end{figure}
\begin{figure}[!h] 
\includegraphics[width=0.80 \linewidth, angle=-0,keepaspectratio=true,clip=true]{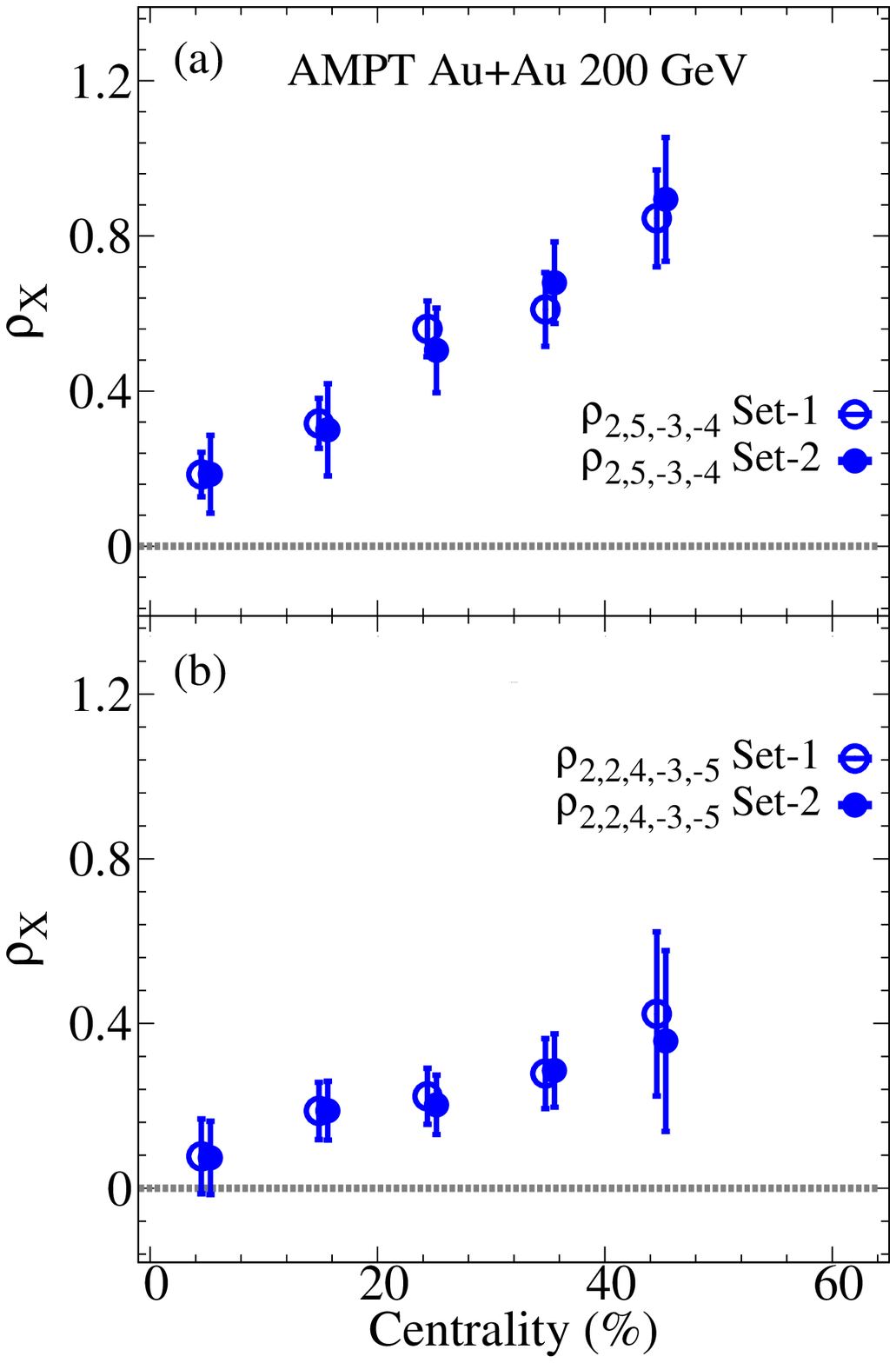}
\vskip -0.4cm
\caption{
The centrality dependence of the normalized asymmetric correlations $\rho_{2,5,-3,-4}$ panel (a) and $\rho_{2,2,4,-3,-5}$ panel (b), for Au+Au at $\sqrt{\textit{s}_{NN}}~=$ 200~GeV from the AMPT model Set-1 and Set-2.
}\label{fig:r5}
\vskip -0.3cm
\end{figure}

Figure~\ref{fig:r3} show the centrality and the $\sigma_{pp}$ dependence of the normalized asymmetric correlations $\rho_{1,1,-2}$ and $\rho_{1,1,1,-1,-2}$ panel (a), $\rho_{2,2,2,-3,-3}$ panel (b), $\rho_{2,2,-4}$, $\rho_{2,2,2,-2,-4}$, and $\rho_{2,2,3,-3,-4}$ panel (c), and $\rho_{2,4,-6}$ panel (d) from the AMPT model Set-1 and Set-2. 
The ratios $\rho_{1,1,-2}$ and $\rho_{1,1,1,-1,-2}$ indicate that the $\psi_{1}$--$\psi_{2}$ correlations (i.e., $\langle \cos(2\psi_{1} - 2\psi_{2}) \rangle$) increase with centrality selections.
Also, the values agreement between $\rho_{1,1,-2}$ and $\rho_{1,1,1,-1,2}$ suggest that $\rho_{1,1,-2}$ and $\rho_{1,1,1,-1,-2}$ ratios significantly reduce the GMC effects. 
In contrast, the absence of correlations between $\psi_{2}$ and $\psi_{3}$ is demonstrated using the ratio  $\rho_{2,2,2,-3,-3}$ in panel (b).
Such an observation suggests that $\psi_{3}$ in the AMPT model is a fluctuation-driven event plane. 
The correlation between $\psi_{2}$ and $\psi_{4}$ ($\langle \cos(4\psi_{2} - 4\psi_{4}) \rangle$) is given by the ratios $\rho_{2,2,-4}$, $\rho_{2,2,2,-2,-4}$, and $\rho_{2,2,3,-3,-4}$. 
A good agreement between the different ratios is observed, reflecting their ability to give the $\langle \cos(4\psi_{2} - 4\psi_{4}) \rangle$. The data-model comparison in panel (c) indicates that my calculations agree well with the experimental data~\cite{STAR:2019zaf}.
Also, positive correlations between $\psi_2$ and $\psi_6$ ($\langle \cos(6\psi_{2} - 6\psi_{6}) \rangle$) are observed in the centrality dependence of $\rho_{2,4,-6}$.

The centrality and the $\sigma_{pp}$ dependence of the normalized asymmetric correlations $\rho_{1,2,-3}$, $\rho_{1,3,-4}$, $\rho_{2,3,-5}$, $\rho_{2,3,2,-2,-5}$, $\rho_{2,3,3,-3,-5}$, $\rho_{3,3,-2,-4}$, and $\rho_{2,3,3,-4,-4}$ are shown in Fig.~\ref{fig:r4}.
The ratios $\rho_{1,2,-3}$ and $\rho_{1,3,-4}$ indicate that the $\langle \cos(1\psi_{1} + 2\psi_{2} - 3\psi_{3}) \rangle$ and $\langle \cos(1\psi_{1} + 3\psi_{2} - 4\psi_{4}) \rangle$ increase with centrality selections. Also, my AMPT calculations showed comparable correlations for $\rho_{1,2,-3}$ and $\rho_{1,3,-4}$.
The correlation between $\psi_{2}$, $\psi_{3}$, and $\psi_{5}$ ($\langle \cos(2\psi_{2} + 3\psi_{3} - 5\psi_{5}) \rangle$) is defined by the ratios $\rho_{2,3,-3}$, $\rho_{2,3,2,-2,-5}$, and $\rho_{2,3,3,-3,-5}$. The AMPT calculations show a good agreement between these different ratios that reflect their capability to give the $\langle \cos(2\psi_{2} + 3\psi_{3} - 5\psi_{5}) \rangle$. In addition, one finds reasonable agreement between the AMPT calculations and the experimental data~\cite{STAR:2019zaf}.
Also, the correlations between $\psi_2$ ,$\psi_3$, and $\psi_4$ ($\langle  \cos(6\psi_{3} - 2\psi_{2} - 4\psi_{4}) \rangle$ and $\langle  \cos(6\psi_{3} + 2\psi_{2} - 8\psi_{4}) \rangle$) are given by the ratios  $\rho_{3,3,-2,-4}$, and $\rho_{2,3,3,-4,-4}$. The presented ratios are shown to give negative values at the presented centrality selections.

Figure~\ref{fig:r5} show the normalized asymmetric correlations $\rho_{2,5,-3,-4}$ panel (a) and $\rho_{2,2,4,-3,-5}$ panel (b) centrality and the $\sigma_{pp}$ dependence from the AMPT model. 
The ratios $\rho_{2,5,-3,-4}$ and $\rho_{2,2,4,-3,-5}$ give the $\psi_{2}$, $\psi_{3}$, $\psi_{4}$ and, $\psi_{4}$ correlations ($\langle  \cos(2\psi_{2} + 5\psi_{5} - 3\psi_{3} - 4\psi_{4}) \rangle$ and $\langle  \cos(4\psi_{2} + 4\psi_{4} - 3\psi_{3} - 5\psi_{5}) \rangle$). Positive correlations that increase with centrality selections are observed in the AMPT calculations for $\rho_{2,5,-3,-4}$ and $\rho_{2,2,4,-3,-5}$.

The presented $\rho_{X}$ in Figs.~\ref{fig:r3},~\ref{fig:r4}, and ~\ref{fig:r5} indicated no sensitivity to the final state effect given by the change of $\sigma_{pp}$ in the AMPT model. Such observations indicate that  $\rho_{X}$  can be used to constrain the initial state angular correlations.

\section{Summary and outlook}\label{sec:4}
In this work, the AMPT model has been used to investigate the  SC, ASC, NSC, and NASC sensitivity to final state effects and reveal their ability to constrain initial and final state effects simultaneously.
First, this work pointed out that non-flow effects can impact the reliability of the current investigation. The non-flow effects are studied using the HIJING model that contains only non-flow correlations. This study indicates that the presented calculations have little (if any) non-flow contributions. 
Second, the AMPT model is used with two final state setups to study the final state effects on SC, ASC, NSC, and NASC. The AMPT calculations indicate that the correlators studied fall into two broad categories: (I)  correlators (SC and ASC) sensitive to the interplay between initial and final state effects. (II) correlators (NSC and NASC) insensitive to the final state effect and sensitive to initial state fluctuations and correlations. 

In summary,  my work suggests that the SC, ASC, NSC, and NASC can be used to provide simultaneous constraints for the initial and final state effects. In addition, one conclude that further detailed experimental measurements of the SC, ASC, NSC, and NASC over a broad range of beam energies and systems size could operate as an additional constraint for theoretical models.
\section*{Acknowledgments}
The author thanks A. Bilandzic,  C. Zhang, S. Bhatta, and E. Racow for the useful discussions and for pointing out important references.
%
%
This research is supported by the US Department of Energy, Office of Nuclear Physics (DOE NP),  under contracts DE-FG02-87ER40331.A008.

\begin{widetext}
\appendix

\section{Non-flow effects}
\label{Hijing}
Long- and short-range non-flow correlations can impact the multi-particle correlations. The long-range non-flow correlations typically involve jets in a dijet event; however, short-range non-flow correlations contain resonance decays, Bose-Einstein correlation, and fragments of individual jets.
Such correlations usually affect a few particles from one or more $\eta$ regions. The non-flow effects can be reduced by correlating particles from two or more pseudorapidity subevents.
Therefore, a detailed examination of the influence of non-flow effects on the presented observables is required. The HIJING model~\cite{Wang:1991hta,Gyulassy:1994ew}, which contains only non-flow, is an ideal testing ground for evaluating the non-flow effects on the $k$-particle correlations, which are the focus of this paper.

\begin{figure}[!h] 
\includegraphics[width=0.9  \linewidth, angle=-0,keepaspectratio=true,clip=true]{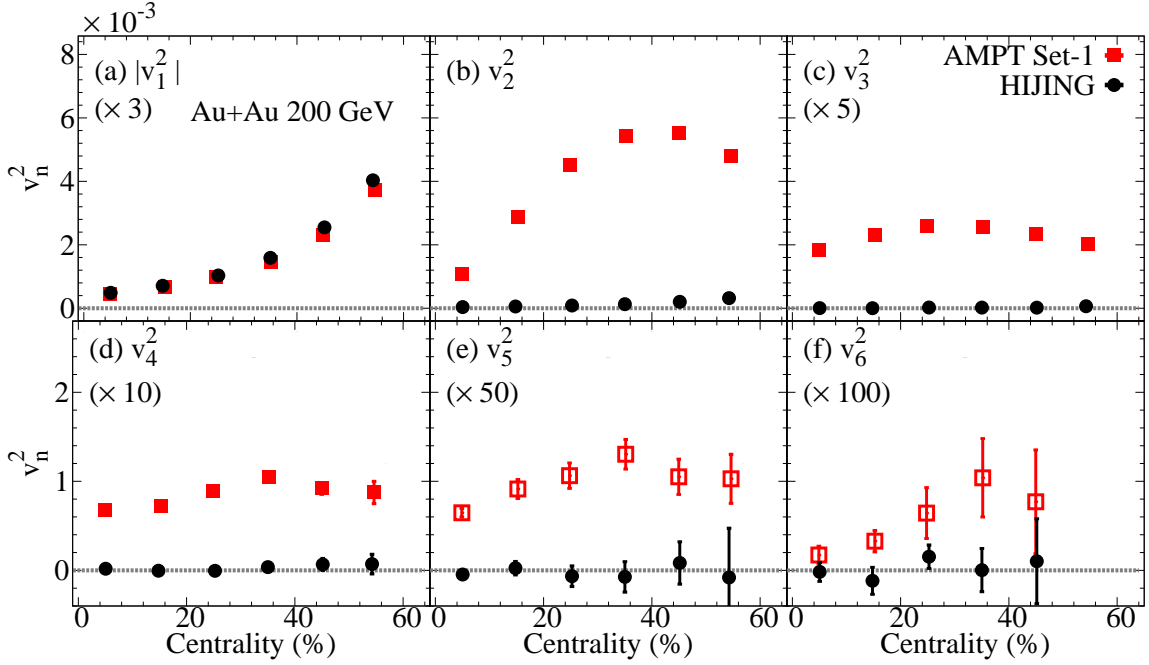}
\vskip -0.4cm
\caption{
The centrality dependence of the two-particle flow harmonics $v^{2}_{n} = SC$($n$,$-n$) using two-subevents method for Au+Au at $\sqrt{\textit{s}_{NN}}~=$ 200~GeV from the AMPT model Set-1 and the HIJING model.
}\label{fig:h1}
\vskip -0.3cm
\end{figure}
\begin{figure}[!h] 
\includegraphics[width=0.9 \linewidth, angle=-0,keepaspectratio=true,clip=true]{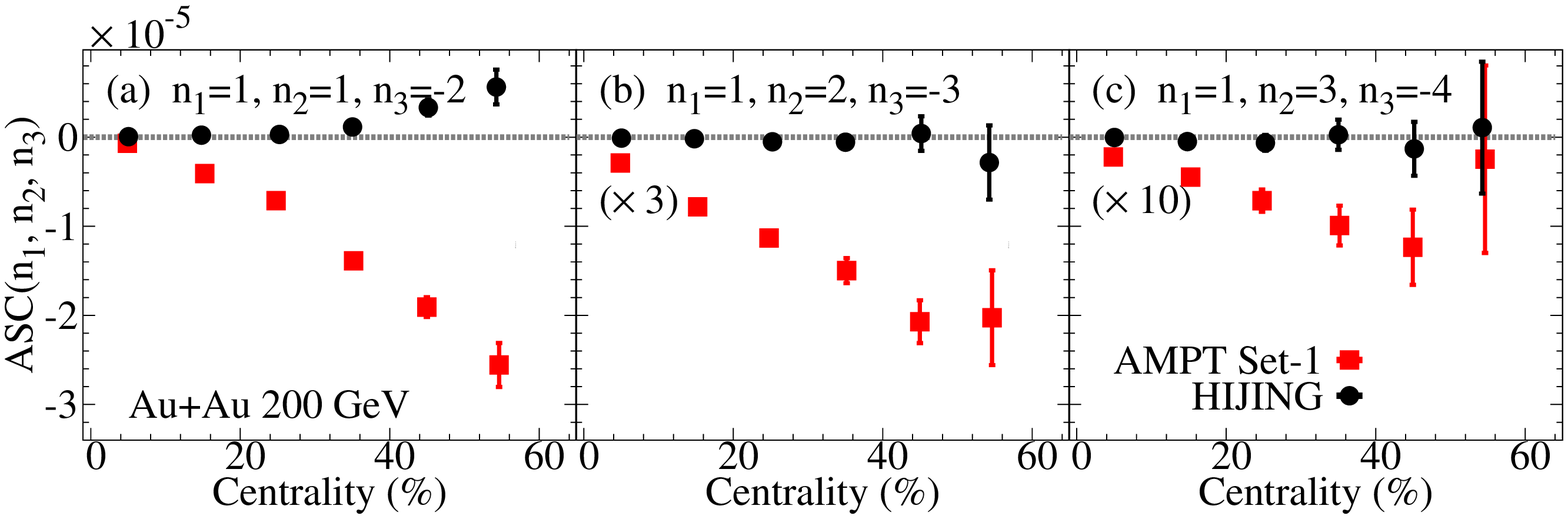}
\vskip -0.4cm
\caption{
The centrality dependence of the three-particle asymmetric correlations $ASC$($1$,$1$,$-2$) panel (a),  $ASC$($1$,$2$,$-3$) panel (b) and  $ASC$($1$,$3$,$-4$) panel (c),  using two-subevents method for Au+Au at $\sqrt{\textit{s}_{NN}}~=$ 200~GeV from the AMPT model Set-1 and the HIJING model.
}\label{fig:h2}
\vskip -0.3cm
\end{figure}
\begin{figure}[!h] 
\includegraphics[width=0.9 \linewidth, angle=-0,keepaspectratio=true,clip=true]{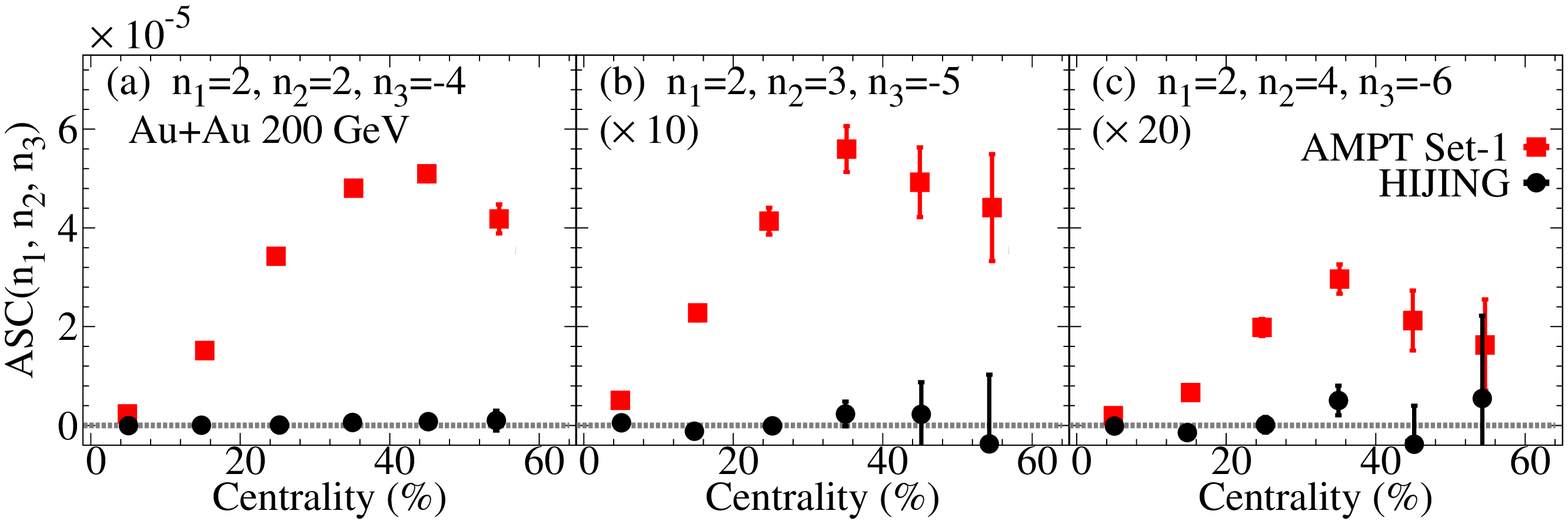}
\vskip -0.4cm
\caption{
The centrality dependence of the three-particle asymmetric correlations $ASC$($2$,$2$,$-4$) panel (a),  $ASC$($2$,$3$,$-5$) panel (b) and  $ASC$($2$,$4$,$-5$) panel (c), using two-subevents method for Au+Au at $\sqrt{\textit{s}_{NN}}~=$ 200~GeV from the AMPT model Set-1 and the HIJING model.
}\label{fig:h3}
\vskip -0.3cm
\end{figure}
\begin{figure}[!h] 
\includegraphics[width=0.9 \linewidth, angle=-0,keepaspectratio=true,clip=true]{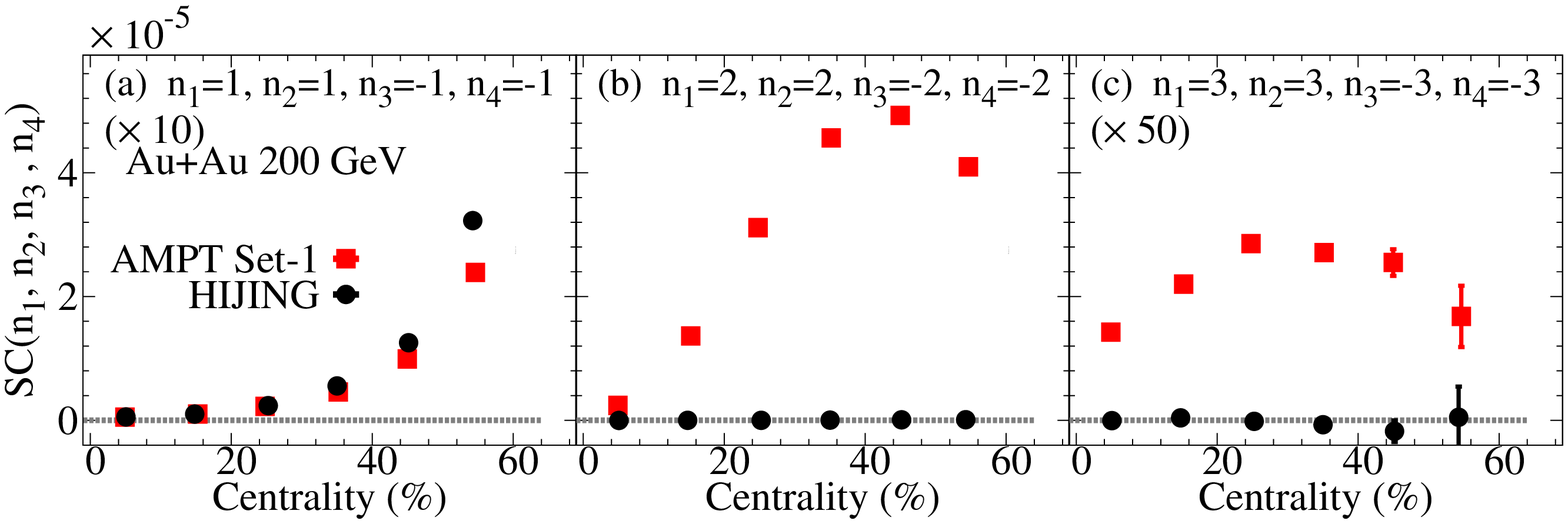}
\vskip -0.4cm
\caption{
The centrality dependence of the four-particle symmetric correlations $SC$($1$,$1$,$-1$,$-1$) panel (a),  $SC$($2$,$2$,$-2$,$-2$) panel (b) and  $SC$($3$,$3$,$-3$,$-3$) panel (c), using two-subevents method for Au+Au at $\sqrt{\textit{s}_{NN}}~=$ 200~GeV from the AMPT model Set-1 and the HIJING model.
}\label{fig:h4}
\vskip -0.3cm
\end{figure}
\begin{figure}[!h] 
\includegraphics[width=0.9 \linewidth, angle=-0,keepaspectratio=true,clip=true]{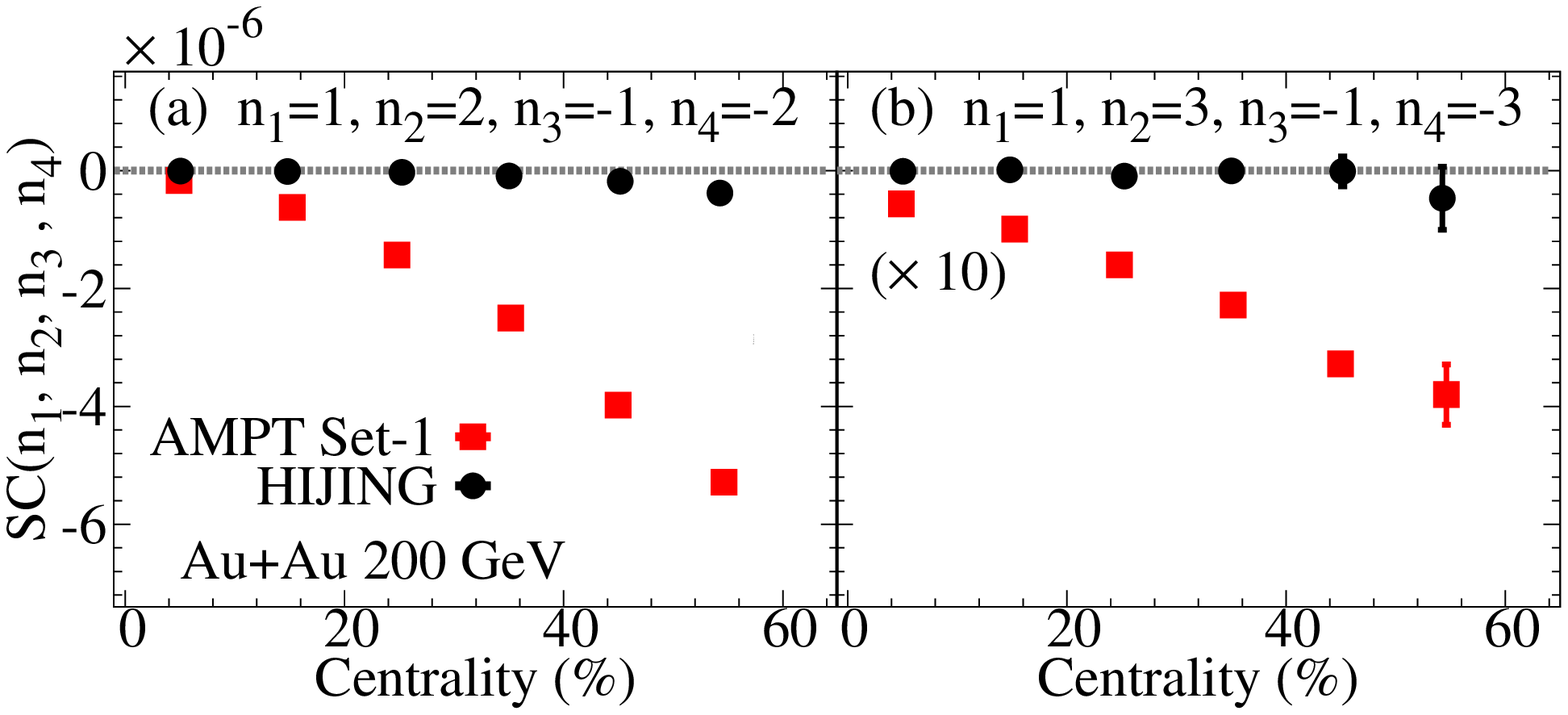}
\vskip -0.4cm
\caption{
The centrality dependence of the four-particle symmetric correlations $SC$($1$,$2$,$-1$,$-2$) panel (a) and $SC$($1$,$3$,$-1$,$-3$) panel (b), using two-subevents method for Au+Au at $\sqrt{\textit{s}_{NN}}~=$ 200~GeV from the AMPT model Set-1 and the HIJING model.
}\label{fig:h5}
\vskip -0.3cm
\end{figure}
\begin{figure}[!h] 
\includegraphics[width=0.9 \linewidth, angle=-0,keepaspectratio=true,clip=true]{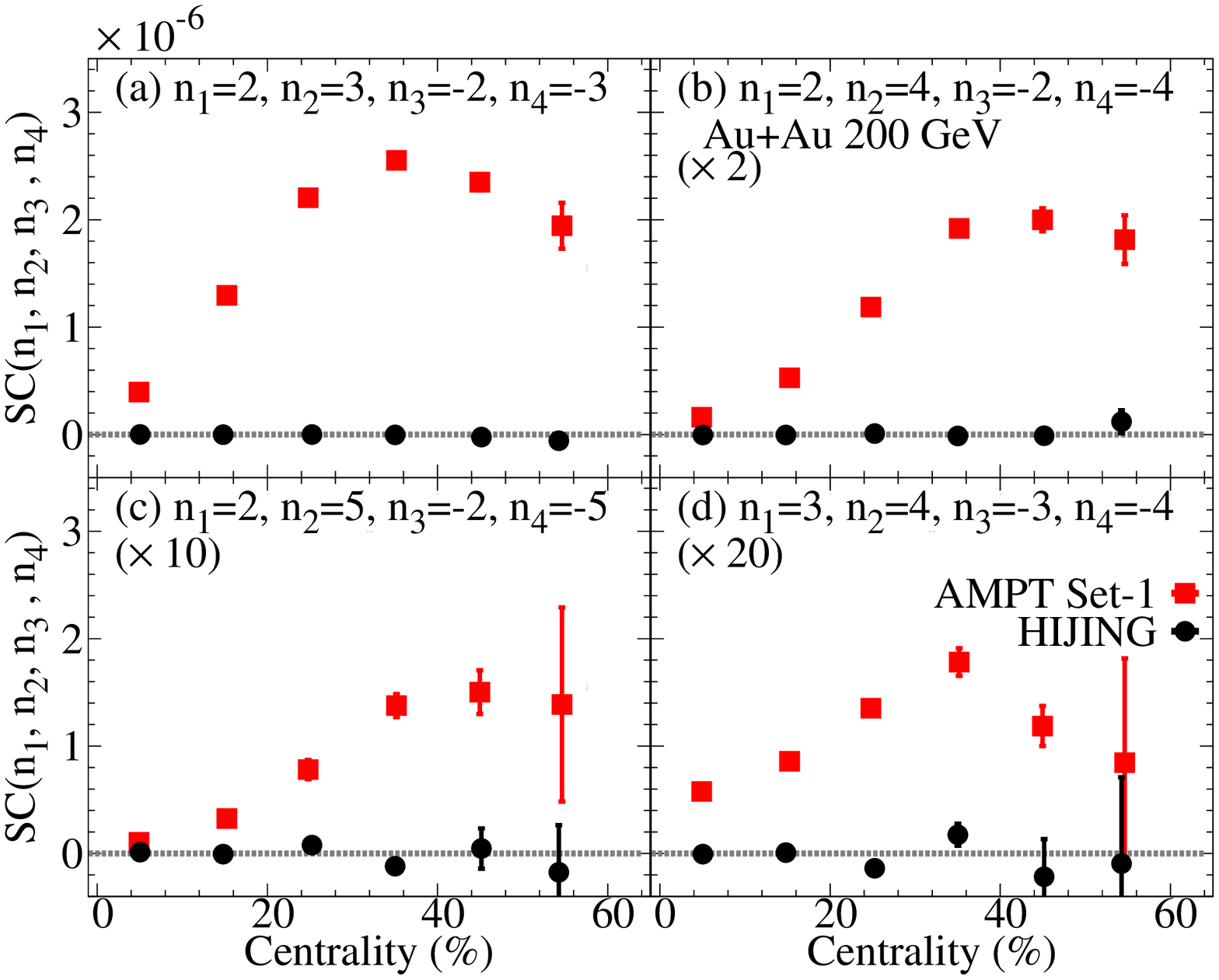}
\vskip -0.4cm
\caption{
The centrality dependence of the four-particle symmetric correlations $SC$($2$,$3$,$-2$,$-3$) panel (a),  $SC$($2$,$4$,$-2$,$-4$) panel (b), $SC$($2$,$4$,$-2$,$-4$) panel (c), and $SC$($3$,$4$,$-3$,$-4$) panel (d), using two-subevents method for Au+Au at $\sqrt{\textit{s}_{NN}}~=$ 200~GeV from the AMPT model Set-1 and the HIJING model.
}\label{fig:h6}
\vskip -0.3cm
\end{figure}
\begin{figure}[!h] 
\includegraphics[width=0.9 \linewidth, angle=-0,keepaspectratio=true,clip=true]{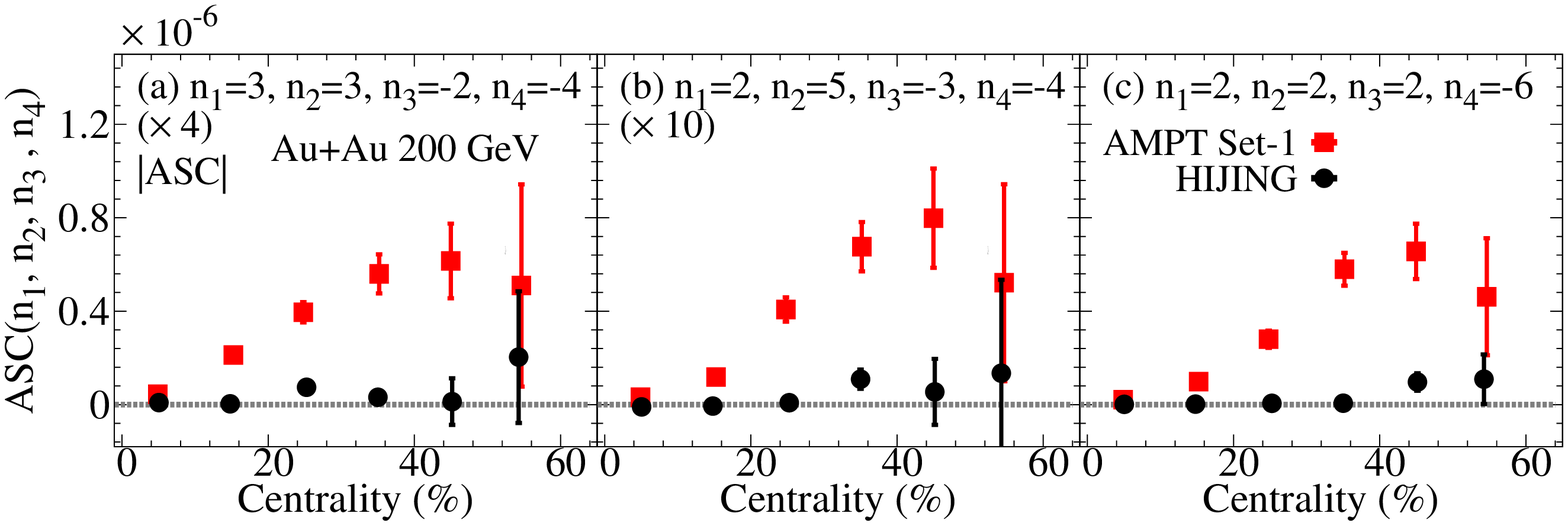}
\vskip -0.4cm
\caption{
The centrality dependence of the four-particle asymmetric correlations $ASC$($3$,$3$,$-2$,$-4$) panel (a),  $ASC$($2$,$5$,$-3$,$-4$) panel (b) and  $ASC$($2$,$2$,$2$,$-6$) panel (c), using two-subevents method for Au+Au at $\sqrt{\textit{s}_{NN}}~=$ 200~GeV from the AMPT model Set-1 and the HIJING model.
}\label{fig:h7}
\vskip -0.3cm
\end{figure}
\begin{figure}[!h] 
\includegraphics[width=0.9 \linewidth, angle=-0,keepaspectratio=true,clip=true]{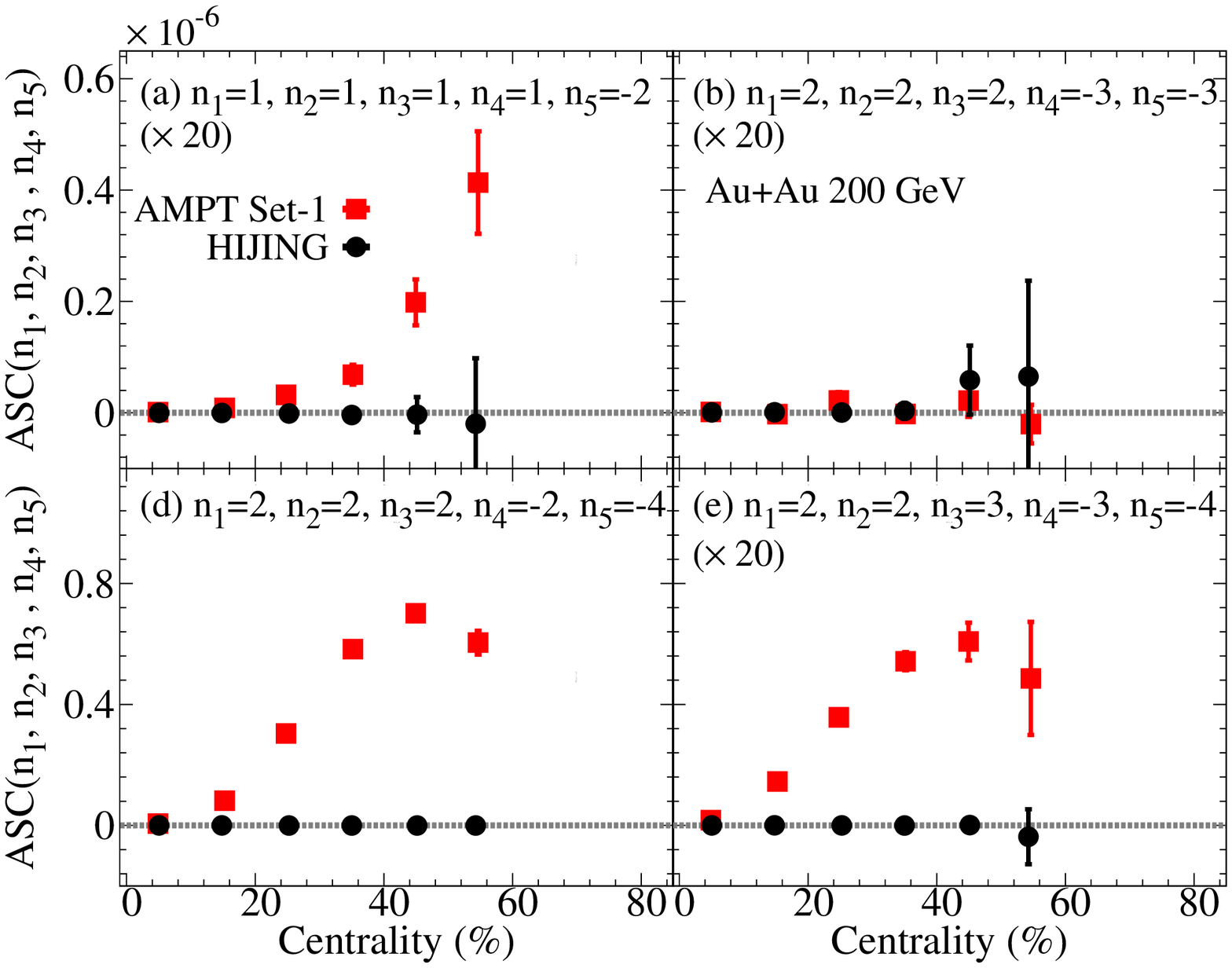}
\vskip -0.4cm
\caption{
The centrality dependence of the five-particle asymmetric correlations $ASC$($1$,$1$,$1$,$-1$,$-2$) panel (a),  $ASC$($2$,$2$,$2$,$-3$,$-3$) panel (b), $ASC$($2$,$2$,$2$,$-2$,$-4$) panel (c), and $ASC$($2$,$2$,$3$,$-3$,$-4$) panel (d), using one-subevent method for Au+Au at $\sqrt{\textit{s}_{NN}}~=$ 200~GeV from the AMPT model Set-1 and the HIJING model.
}\label{fig:h8}
\vskip -0.3cm
\end{figure}
\begin{figure}[!h] 
\includegraphics[width=0.9 \linewidth, angle=-0,keepaspectratio=true,clip=true]{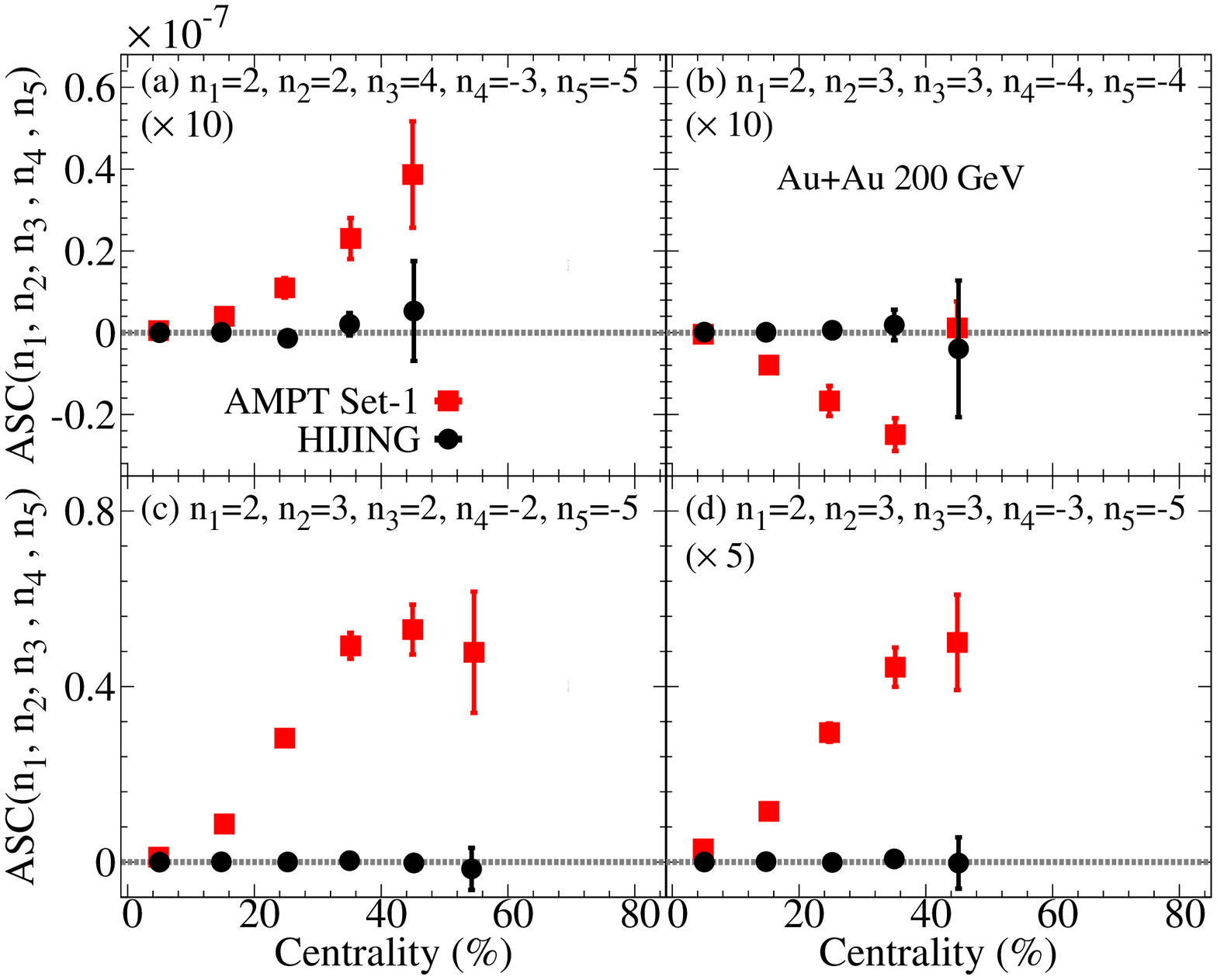}
\vskip -0.4cm
\caption{
The centrality dependence of the five-particle asymmetric correlations $ASC$($2$,$2$,$4$,$-3$,$-5$) panel (a),  $ASC$($2$,$3$,$3$,$-4$,$-4$) panel (b), $ASC$($2$,$3$,$2$,$-2$,$-5$) panel (c), and $ASC$($2$,$3$,$3$,$-3$,$-5$) panel (d), using one-subevent method for Au+Au at $\sqrt{\textit{s}_{NN}}~=$ 200~GeV from the AMPT model Set-1 and the HIJING model.
}\label{fig:h9}
\vskip -0.3cm
\end{figure}
\begin{figure}[!h] 
\includegraphics[width=0.9 \linewidth, angle=-0,keepaspectratio=true,clip=true]{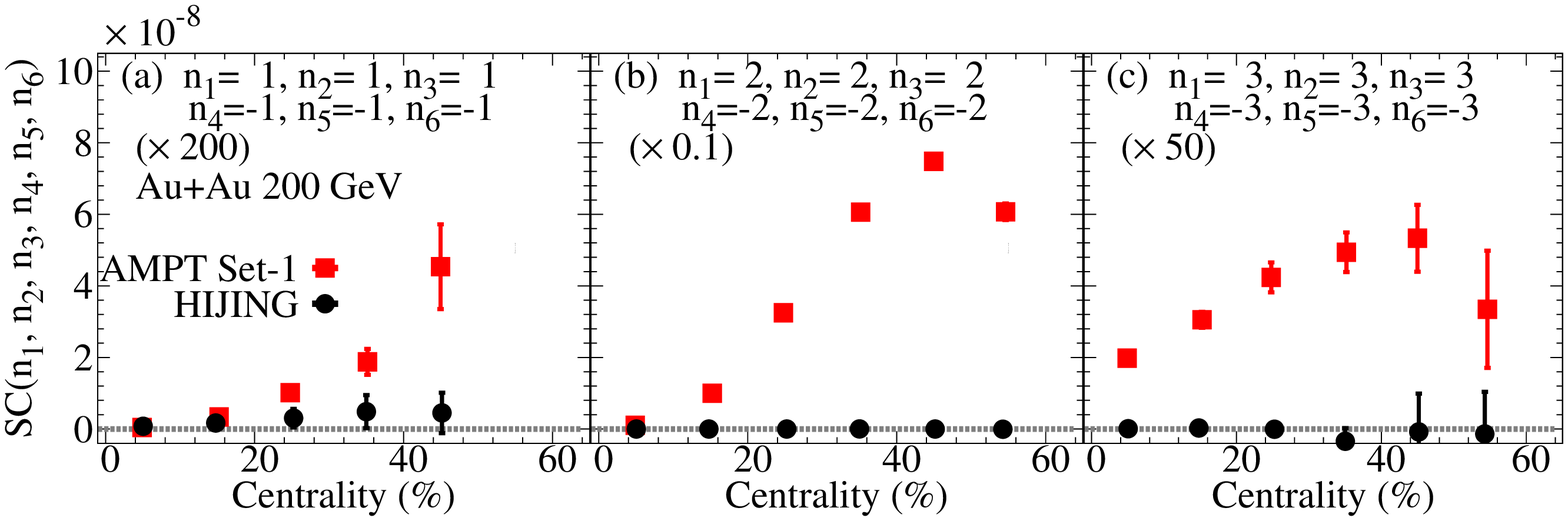}
\vskip -0.4cm
\caption{
The centrality dependence of the six-particle symmetric correlations $SC$($1$,$1$,$1$,$-1$,$-1$,$-2$) panel (a),  $SC$($2$,$2$,$2$,$-2$,$-2$,$-2$) panel (b) and  $SC$($3$,$3$,$3$,$-3$,$-3$,$-3$) panel (c), using one-subevent method for Au+Au at $\sqrt{\textit{s}_{NN}}~=$ 200~GeV from the AMPT model Set-1 and the HIJING model.
}\label{fig:h101}
\vskip -0.3cm
\end{figure}
\begin{figure}[!h] 
\includegraphics[width=0.9 \linewidth, angle=-0,keepaspectratio=true,clip=true]{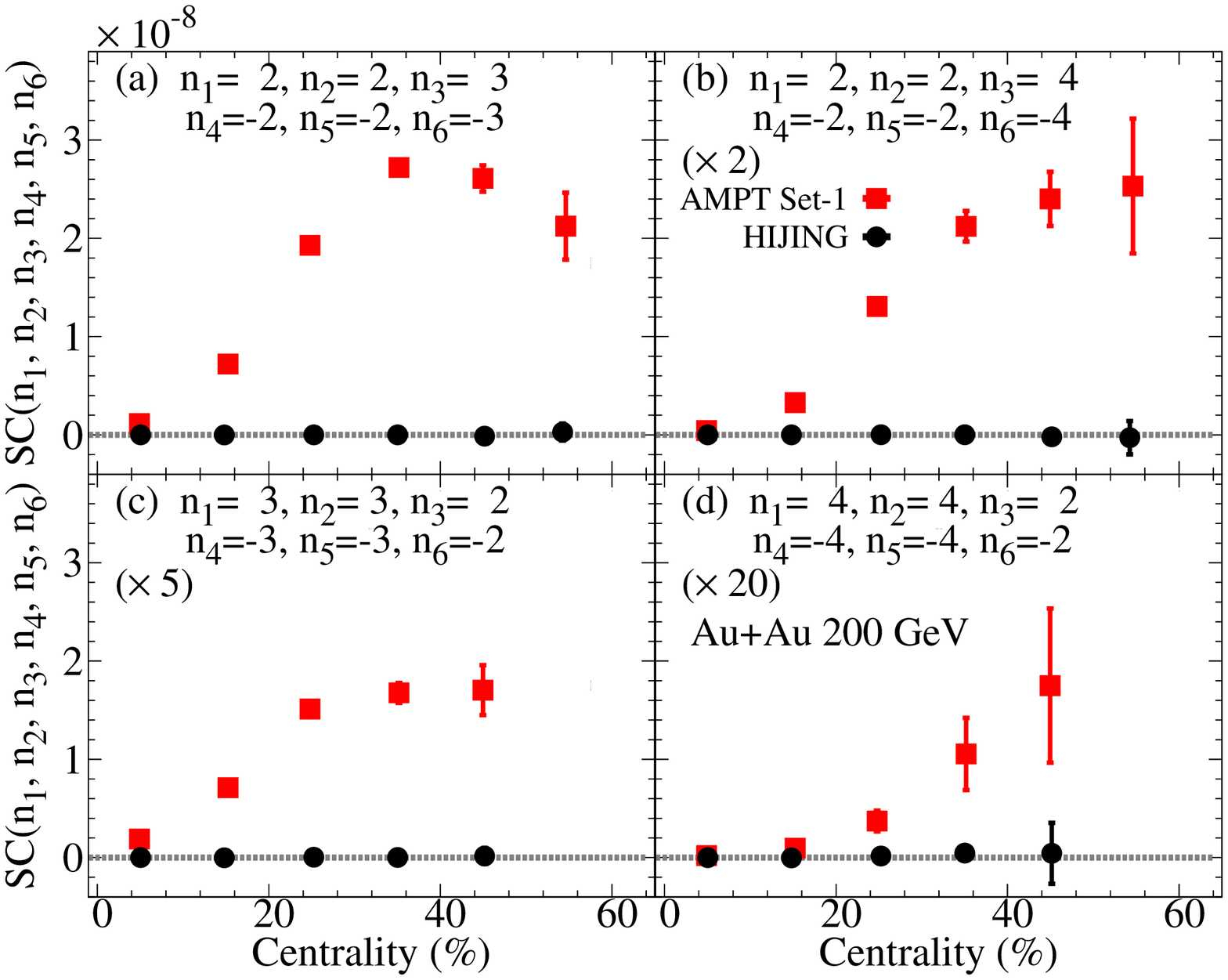}
\vskip -0.4cm
\caption{
The centrality dependence of the six-particle symmetric correlations $SC$($2$,$2$,$3$,$-2$,$-2$,$-3$) panel (a),  $SC$($2$,$2$,$4$,$-2$,$-2$,$-4$) panel (b),  $SC$($3$,$3$,$2$,$-3$,$-3$,$-2$) panel (c), and $SC$($4$,$4$,$2$,$-4$,$-4$,$-2$) panel (d),  using one-subevent method for Au+Au at $\sqrt{\textit{s}_{NN}}~=$ 200~GeV from the AMPT model Set-1 and the HIJING model.
}\label{fig:h102}
\vskip -0.3cm
\end{figure}
\begin{figure}[!h] 
\includegraphics[width=0.9 \linewidth, angle=-0,keepaspectratio=true,clip=true]{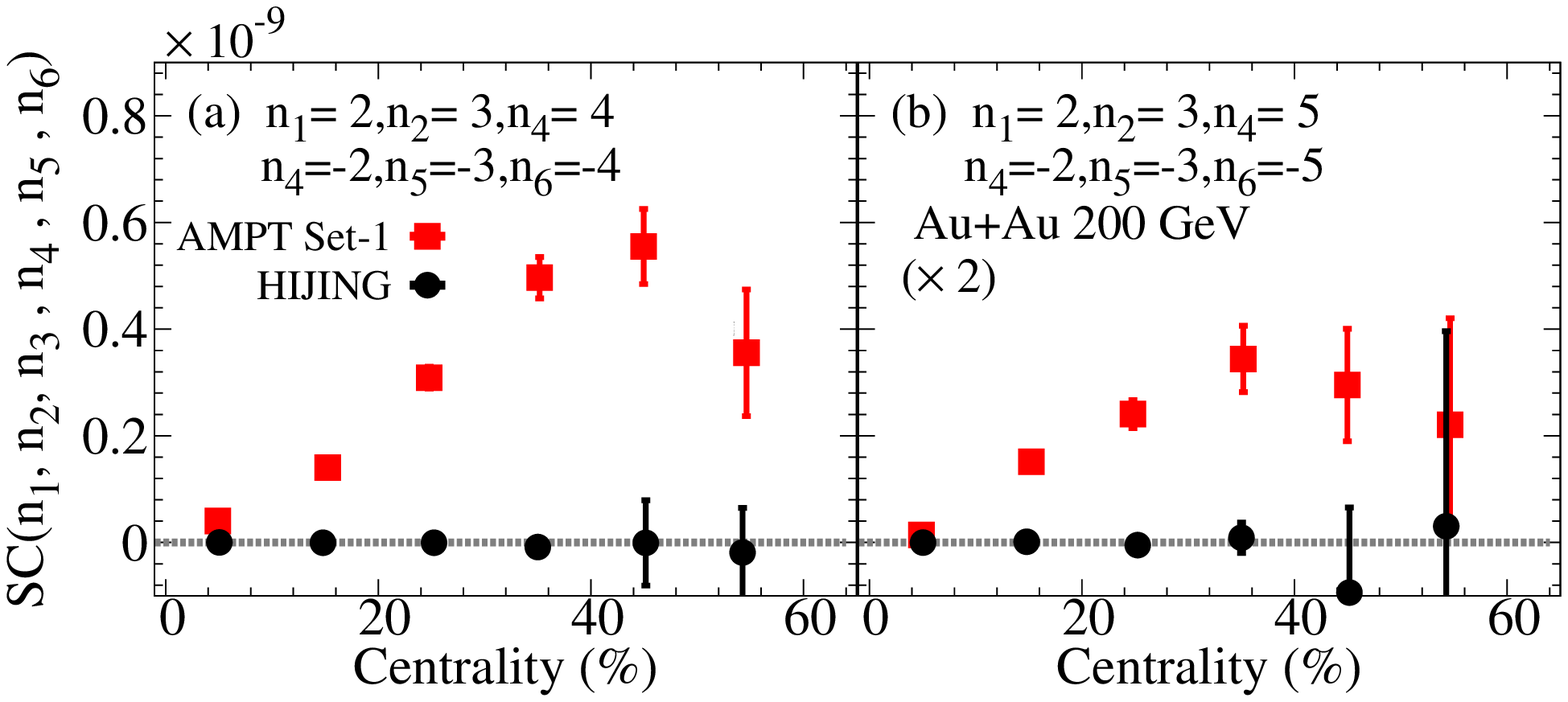}
\vskip -0.4cm
\caption{
The centrality dependence of the six-particle symmetric correlations $SC$($2$,$3$,$4$,$-2$,$-3$,$-4$) panel (a) and  $SC$($2$,$3$,$5$,$-2$,$-3$,$-5$) panel (b), using one-subevent method for Au+Au at $\sqrt{\textit{s}_{NN}}~=$ 200~GeV from the AMPT model Set-1 and the HIJING model.
}\label{fig:h103}
\vskip -0.3cm
\end{figure}
 Figures~\ref{fig:h1}--\ref{fig:h103} show that using the two-subevents method substantial reduce the non flow effects on the two-, three-, and four-particles correlations. The main intention of this HIJING investigation is to compare the order of magnitude between the signal given in the AMPT model vs. what correlations in the HIJING model. These results Figs.~\ref{fig:h1}--\ref{fig:h103} are compatible with other investigations~\cite{Huo:2017nms,Zhang:2021phk}.


\section{Formulas}\label{cum}
The events generated were analyzed using the  two- and multi-particle correlations given via the use of the subevents cumulant methods~\cite{Jia:2017hbm,Huo:2017nms,Zhang:2018lls,Magdy:2020bhd}. The observables discussed in this work can be given in terms of the flow vectors $Q_{n}$ as;
\begin{eqnarray}\label{eq:A2-1}
Q(n,k)              &=&  \sum^{M}_{i=1} \omega^{k}_{i} e^{in\varphi_{i}},
\end{eqnarray}
where $\phi_i$ is the azimuthal angle of the $\mathit{i}^{th}$ particle. The order $n$ can take positive or negative values, for $n$=0 Eq.~\ref{eq:A2-1} $Q(0,k)$ gives the event multiplicity. 

\begin{itemize}
\item{The one-subevents two-particle $SC(n_1,n_2)$;}
\end{itemize}

{\small \begin{eqnarray}\label{eq:A2-1}
SC(n_1,n_2)  &=& \langle Two(n_1,n_2) \rangle / \langle Two(0,0) \rangle \\  \nonumber
Two(n_1,n_2) &=&  Q(n_1,1) Q(n_2,1) - Q(n_1+n_2,2)
\end{eqnarray} }

\begin{itemize}
\item{The one-subevents three-particle $ASC(n_1,n_2,n_3)$;}
\end{itemize}

{\small \begin{eqnarray}\label{eq:A2-1}
ASC(n_1,n_2,n_3)  &=& \langle Three(n_1,n_2,n_3) \rangle / \langle Three(0,0,0) \rangle \\  \nonumber
Three(n_1,n_2,n_3) &=&  Q(n_1,1) Q(n_2,1) Q(n_3,1) \\  \nonumber
                &-&  Q(n_1+n_2,2) Q(n_3,1) - Q(n_1+n_3,2) Q(n_2,1) - Q(n_2+n_3,2) Q(n_1,1) \\  \nonumber
                &+&  2 Q(n_1+n_2+n_3,3)
\end{eqnarray} }

\begin{itemize}
\item{The one-subevents four-particle $SC(n_1,n_2,n_3,n_4)$ and $ASC(n_1,n_2,n_3,n_4)$;}
\end{itemize}

{\small  \begin{eqnarray}\label{eq:A2-1}
ASC(n_1,n_2,n_3,n_4)  &=& \langle Four(n_1,n_2,n_3,n_4) \rangle / \langle Four(0,0,0,0) \rangle \\  
Four(n_1,n_2,n_3,n_4) &=& Q(n_1,1) Q(n_2,1) Q(n_3,1) Q(n_4,1) \\  \nonumber
                      &-& Q(n_1+n_2,2) Q(n_3,1) Q(n_4,1) - Q(n_1+n_3,2) Q(n_2,1) Q(n_4,1) \\  \nonumber
                      &-& Q(n_1+n_4,2) Q(n_2,1) Q(n_3,1) - Q(n_2+n_3,2) Q(n_1,1) Q(n_4,1) \\  \nonumber
                      &-& Q(n_2+n_4,2) Q(n_1,1) Q(n_3,1) - Q(n_3+n_4,2) Q(n_1,1) Q(n_2,1) \\  \nonumber
                      &+& 2 Q(n_1+n_2+n_3,3) Q(n_4,1)    + 2 Q(n_1+n_2+n_4,3) Q(n_3,1) \\  \nonumber
                      &+& 2 Q(n_1+n_3+n_4,3) Q(n_2,1)    + 2 Q(n_2+n_3+n_4,3) Q(n_1,1) \\  \nonumber
                      &+& Q(n_1+n_2,2) Q(n_3+n_4,2) + Q(n_1+n_3,2) Q(n_2+n_4,2) + Q(n_2+n_3,2) Q(n_1+n_4,2) \\  \nonumber
                      &-& 6 Q(n_1+n_2+n_3+n_4,4). \\  \nonumber
\end{eqnarray} }

\begin{itemize}
\item{The one-subevents five-particle  $ASC(n_1,n_2,n_3,n_4,n_5)$;}
\end{itemize}
{\small
\begin{eqnarray}\label{eq:A2-1}
ASC(n_1,n_2,n_3,n_4,n_5)  &=& \langle Five(n_1,n_2,n_3,n_4,n_5) \rangle / \langle Five(0,0,0,0,0) \rangle \\ 
Five(n_1,n_2,n_3,n_4,n_5) &=& A1 + A2 + A3 + A4 + A5 + A6 + A7,     \\  \nonumber
\end{eqnarray}
\begin{eqnarray}
    A1     &=&   Q(n_1,1) Q(n_2,1) Q(n_3,1) Q(n_4,1) Q(n_5,1),      \\  \nonumber
\end{eqnarray}
\begin{eqnarray}
    A2     &=& - Q(n_1+n_2,2) Q(n_3,1) Q(n_4,1) Q(n_5,1) - Q(n_1+n_3,2) Q(n_2,1) Q(n_4,1) Q(n_5,1)  \\  \nonumber
           &-& Q(n_1+n_4,2) Q(n_2,1) Q(n_3,1) Q(n_5,1) - Q(n_1+n_5,2) Q(n_2,1) Q(n_3,1) Q(n_4,1)     \\  \nonumber
           &-& Q(n_2+n_3,2) Q(n_1,1) Q(n_4,1) Q(n_5,1) - Q(n_2+n_4,2) Q(n_1,1) Q(n_3,1) Q(n_5,1)     \\  \nonumber
           &-& Q(n_2+n_5,2) Q(n_1,1) Q(n_3,1) Q(n_4,1) - Q(n_3+n_4,2) Q(n_1,1) Q(n_2,1) Q(n_5,1)     \\  \nonumber
           &-& Q(n_3+n_5,2) Q(n_1,1) Q(n_2,1) Q(n_4,1) - Q(n_4+n_5,2) Q(n_1,1) Q(n_2,1) Q(n_3,1),     \\  \nonumber
\end{eqnarray}
\begin{eqnarray}
    A3     &=& 2 Q(n_1+n_2+n_3,3) Q(n_4,1) Q(n_5,1) + 2 Q(n_1+n_2+n_4,3) Q(n_3,1) Q(n_5,1)  \\  \nonumber
           &+& 2 Q(n_1+n_2+n_5,3) Q(n_3,1) Q(n_4,1) + 2 Q(n_1+n_3+n_4,3) Q(n_2,1) Q(n_5,1)  \\  \nonumber
           &+& 2 Q(n_1+n_3+n_5,3) Q(n_2,1) Q(n_4,1) + 2 Q(n_1+n_4+n_5,3) Q(n_2,1) Q(n_3,1)  \\  \nonumber
           &+& 2 Q(n_2+n_3+n_4,3) Q(n_1,1) Q(n_5,1) + 2 Q(n_2+n_3+n_5,3) Q(n_1,1) Q(n_4,1)  \\  \nonumber
           &+& 2 Q(n_2+n_4+n_5,3) Q(n_1,1) Q(n_3,1) + 2 Q(n_3+n_4+n_5,3) Q(n_1,1) Q(n_2,1), \\  \nonumber
\end{eqnarray}
\begin{eqnarray}
    A4     &=& Q(n_1+n_2,2) Q(n_3+n_4,2) Q(n_5,1) + Q(n_1+n_2,2) Q(n_3+n_5,2) Q(n_4,1)  \\  \nonumber
           &+& Q(n_1+n_2,2) Q(n_4+n_5,2) Q(n_3,1) + Q(n_1+n_3,2) Q(n_2+n_4,2) Q(n_5,1)  \\  \nonumber
           &+& Q(n_1+n_3,2) Q(n_2+n_5,2) Q(n_4,1) + Q(n_1+n_3,2) Q(n_4+n_5,2) Q(n_2,1)  \\  \nonumber
           &+& Q(n_1+n_4,2) Q(n_2+n_3,2) Q(n_5,1) + Q(n_1+n_4,2) Q(n_2+n_5,2) Q(n_3,1)  \\  \nonumber
           &+& Q(n_1+n_4,2) Q(n_3+n_5,2) Q(n_2,1) + Q(n_1+n_5,2) Q(n_2+n_3,2) Q(n_4,1)  \\  \nonumber
           &+& Q(n_1+n_5,2) Q(n_2+n_4,2) Q(n_3,1) + Q(n_1+n_5,2) Q(n_3+n_4,2) Q(n_2,1)  \\  \nonumber
           &+& Q(n_2+n_5,2) Q(n_3+n_4,2) Q(n_1,1) + Q(n_2+n_3,2) Q(n_4+n_5,2) Q(n_1,1)  \\  \nonumber
           &+& Q(n_2+n_4,2) Q(n_3+n_5,2) Q(n_1,1),      \\  \nonumber
\end{eqnarray}
\begin{eqnarray}
    A5     &=& - 2 Q(n_1+n_2+n_3,3) Q(n_4+n_5,2) - 2 Q(n_1+n_2+n_4,3) Q(n_3+n_5,2) \\  \nonumber
           &-&   2 Q(n_1+n_2+n_5,3) Q(n_3+n_4,2) - 2 Q(n_1+n_3+n_4,3) Q(n_2+n_5,2)   \\  \nonumber
           &-&   2 Q(n_1+n_3+n_5,3) Q(n_2+n_4,2) - 2 Q(n_1+n_4+n_5,3) Q(n_2+n_3,2)   \\  \nonumber
           &-&   2 Q(n_2+n_3+n_4,3) Q(n_1+n_5,2) - 2 Q(n_2+n_3+n_5,3) Q(n_1+n_4,2)   \\  \nonumber
           &-&   2 Q(n_2+n_4+n_5,3) Q(n_1+n_3,2) - 2 Q(n_3+n_4+n_5,3) Q(n_1+n_2,2),   \\  \nonumber
\end{eqnarray}
\begin{eqnarray}
    A6     &=& - 6 Q(n_1+n_2+n_3+n_4,4) Q(n_5,1) - 6 Q(n_1+n_2+n_3+n_5,4) Q(n_4,1)  \\  \nonumber
           &-&   6 Q(n_1+n_2+n_4+n_5,4) Q(n_3,1) - 6 Q(n_1+n_3+n_4+n_5,4) Q(n_2,1)  \\  \nonumber
           &-&   6 Q(n_2+n_3+n_4+n_5,4) Q(n_1,1),      \\  \nonumber
\end{eqnarray}
\begin{eqnarray}
    A7     &=&  24 Q(n_1+n_2+n_3+n_4+n_5,5).     \\  \nonumber
\end{eqnarray}

}

\begin{itemize}
\item{The one-subevents six-particle  $SC(n_1,n_2,n_3,n_4,n_5,n_6)$;}
\end{itemize}
{\small
\begin{eqnarray}\label{eq:A2-1}
SC(n_1,n_2,n_3,n_4,n_5,n_6)  &=& \langle Six(n_1,n_2,n_3,n_4,n_5,n_6) \rangle / \langle Six(0,0,0,0,0,0) \rangle \\ 
Six(n_1,n_2,n_3,n_4,n_5,n_6) &=& B1 + B2 + B3 + B4 + B5 + B6 + B7 + B8 + B9 + B10 + B11,     \\  \nonumber
\end{eqnarray}
\begin{eqnarray}
B1 &=&  Q(n_1,1) Q(n_2,1) Q(n_3,1) Q(n_4,1) Q(n_5,1) Q(n_6,1),    \\  \nonumber
\end{eqnarray}
\begin{eqnarray}
B2 &=&     \\  \nonumber
     &-& Q(n_1,1) Q(n_2,1) Q(n_3,1) Q(n_4,1) Q(n_5+n_6,2) - Q(n_1,1) Q(n_2,1) Q(n_3,1) Q(n_5,1) Q(n_4+n_6,2)   \\  \nonumber
     &-& Q(n_1,1) Q(n_2,1) Q(n_4,1) Q(n_5,1) Q(n_3+n_6,2) - Q(n_1,1) Q(n_3,1) Q(n_4,1) Q(n_5,1) Q(n_2+n_6,2)   \\  \nonumber
     &-& Q(n_3,1) Q(n_4,1) Q(n_5,1) Q(n_6,1) Q(n_1+n_2,2) - Q(n_2,1) Q(n_4,1) Q(n_5,1) Q(n_6,1) Q(n_1+n_3,2)   \\  \nonumber
     &-& Q(n_1,1) Q(n_4,1) Q(n_5,1) Q(n_6,1) Q(n_2+n_3,2) - Q(n_2,1) Q(n_3,1) Q(n_5,1) Q(n_6,1) Q(n_1+n_4,2)   \\  \nonumber
     &-& Q(n_1,1) Q(n_3,1) Q(n_5,1) Q(n_6,1) Q(n_2+n_4,2) - Q(n_1,1) Q(n_2,1) Q(n_5,1) Q(n_6,1) Q(n_3+n_4,2)   \\  \nonumber
     &-& Q(n_2,1) Q(n_3,1) Q(n_4,1) Q(n_6,1) Q(n_1+n_5,2) - Q(n_1,1) Q(n_3,1) Q(n_4,1) Q(n_6,1) Q(n_2+n_5,2)   \\  \nonumber
     &-& Q(n_1,1) Q(n_2,1) Q(n_4,1) Q(n_6,1) Q(n_3+n_5,2) - Q(n_1,1) Q(n_2,1) Q(n_3,1) Q(n_6,1) Q(n_4+n_5,2)   \\  \nonumber
     &-& Q(n_2,1) Q(n_3,1) Q(n_4,1) Q(n_5,1) Q(n_1+n_6,2),
\end{eqnarray}
\begin{eqnarray} 
B3 &=&     \\  \nonumber
      &~& 2 Q(n_1+n_2+n_3,3) Q(n_4,1) Q(n_5,1) Q(n_6,1) + 2 Q(n_2,1) Q(n_1+n_3+n_4,3) Q(n_5,1) Q(n_6,1)   \\  \nonumber
      &+& 2 Q(n_1,1) Q(n_2+n_3+n_4,3) Q(n_5,1) Q(n_6,1) + 2 Q(n_3,1) Q(n_1+n_2+n_4,3) Q(n_5,1) Q(n_6,1)   \\  \nonumber
      &+& 2 Q(n_3,1) Q(n_4,1) Q(n_1+n_2+n_5,3) Q(n_6,1) + 2 Q(n_2,1) Q(n_4,1) Q(n_1+n_3+n_5,3) Q(n_6,1)   \\  \nonumber
      &+& 2 Q(n_1,1) Q(n_4,1) Q(n_2+n_3+n_5,3) Q(n_6,1) + 2 Q(n_2,1) Q(n_3,1) Q(n_1+n_4+n_5,3) Q(n_6,1)   \\  \nonumber
      &+& 2 Q(n_1,1) Q(n_3,1) Q(n_2+n_4+n_5,3) Q(n_6,1) + 2 Q(n_1,1) Q(n_2,1) Q(n_3+n_4+n_5,3) Q(n_6,1)   \\  \nonumber
      &+& 2 Q(n_3,1) Q(n_4,1) Q(n_5,1) Q(n_1+n_2+n_6,3) + 2 Q(n_2,1) Q(n_4,1) Q(n_5,1) Q(n_1+n_3+n_6,3)   \\  \nonumber
      &+& 2 Q(n_1,1) Q(n_4,1) Q(n_5,1) Q(n_2+n_3+n_6,3) + 2 Q(n_1,1) Q(n_3,1) Q(n_5,1) Q(n_2+n_4+n_6,3)   \\  \nonumber
      &+& 2 Q(n_2,1) Q(n_3,1) Q(n_5,1) Q(n_1+n_4+n_6,3) + 2 Q(n_1,1) Q(n_2,1) Q(n_5,1) Q(n_3+n_4+n_6,3)   \\  \nonumber
      &+& 2 Q(n_2,1) Q(n_3,1) Q(n_4,1) Q(n_1+n_5+n_6,3) + 2 Q(n_1,1) Q(n_3,1) Q(n_4,1) Q(n_2+n_5+n_6,3)   \\  \nonumber
      &+& 2 Q(n_1,1) Q(n_2,1) Q(n_3,1) Q(n_4+n_5+n_6,3) + 2 Q(n_1,1) Q(n_2,1) Q(n_4,1) Q(n_3+n_5+n_6,3),  \\  \nonumber
\end{eqnarray}
\begin{eqnarray}
B4 &=& 
           Q(n_2+n_3,2) Q(n_1+n_4,2) Q(n_5,1) Q(n_6,1) + Q(n_1+n_3,2) Q(n_2+n_4,2) Q(n_5,1) Q(n_6,1)    \\  \nonumber
       &+& Q(n_1+n_2,2) Q(n_3+n_4,2) Q(n_5,1) Q(n_6,1) + Q(n_2+n_3,2) Q(n_4,1) Q(n_1+n_5,2) Q(n_6,1)    \\  \nonumber
       &+& Q(n_3,1) Q(n_2+n_4,2) Q(n_1+n_5,2) Q(n_6,1) + Q(n_2,1) Q(n_3+n_4,2) Q(n_1+n_5,2) Q(n_6,1)    \\  \nonumber
       &+& Q(n_1+n_3,2) Q(n_4,1) Q(n_2+n_5,2) Q(n_6,1) + Q(n_3,1) Q(n_1+n_4,2) Q(n_2+n_5,2) Q(n_6,1)    \\  \nonumber
       &+& Q(n_1,1) Q(n_3+n_4,2) Q(n_2+n_5,2) Q(n_6,1) + Q(n_1+n_2,2) Q(n_4,1) Q(n_3+n_5,2) Q(n_6,1)    \\  \nonumber
       &+& Q(n_2,1) Q(n_1+n_4,2) Q(n_3+n_5,2) Q(n_6,1) + Q(n_1,1) Q(n_2+n_4,2) Q(n_3+n_5,2) Q(n_6,1)    \\  \nonumber
       &+& Q(n_1+n_2,2) Q(n_3,1) Q(n_4+n_5,2) Q(n_6,1) + Q(n_2,1) Q(n_1+n_3,2) Q(n_4+n_5,2) Q(n_6,1)    \\  \nonumber
       &+& Q(n_1,1) Q(n_2+n_3,2) Q(n_4+n_5,2) Q(n_6,1) + Q(n_2+n_3,2) Q(n_4,1) Q(n_5,1) Q(n_1+n_6,2)    \\  \nonumber
       &+& Q(n_3,1) Q(n_2+n_4,2) Q(n_5,1) Q(n_1+n_6,2) + Q(n_2,1) Q(n_3+n_4,2) Q(n_5,1) Q(n_1+n_6,2)    \\  \nonumber
       &+& Q(n_3,1) Q(n_4,1) Q(n_2+n_5,2) Q(n_1+n_6,2) + Q(n_2,1) Q(n_4,1) Q(n_3+n_5,2) Q(n_1+n_6,2)    \\  \nonumber
       &+& Q(n_2,1) Q(n_3,1) Q(n_4+n_5,2) Q(n_1+n_6,2) + Q(n_1+n_3,2) Q(n_4,1) Q(n_5,1) Q(n_2+n_6,2)    \\  \nonumber
       &+& Q(n_3,1) Q(n_1+n_4,2) Q(n_5,1) Q(n_2+n_6,2) + Q(n_1,1) Q(n_3+n_4,2) Q(n_5,1) Q(n_2+n_6,2)    \\  \nonumber
       &+& Q(n_3,1) Q(n_4,1) Q(n_1+n_5,2) Q(n_2+n_6,2) + Q(n_1,1) Q(n_4,1) Q(n_3+n_5,2) Q(n_2+n_6,2)    \\  \nonumber
       &+& Q(n_1+n_2,2) Q(n_4,1) Q(n_5,1) Q(n_3+n_6,2) + Q(n_2,1) Q(n_1+n_4,2) Q(n_5,1) Q(n_3+n_6,2)    \\  \nonumber
       &+& Q(n_1,1) Q(n_2+n_4,2) Q(n_5,1) Q(n_3+n_6,2) + Q(n_2,1) Q(n_4,1) Q(n_1+n_5,2) Q(n_3+n_6,2)    \\  \nonumber
       &+& Q(n_1,1) Q(n_4,1) Q(n_2+n_5,2) Q(n_3+n_6,2) + Q(n_1,1) Q(n_2,1) Q(n_4+n_5,2) Q(n_3+n_6,2)    \\  \nonumber
       &+& Q(n_1,1) Q(n_3,1) Q(n_4+n_5,2) Q(n_2+n_6,2) + Q(n_1+n_2,2) Q(n_3,1) Q(n_5,1) Q(n_4+n_6,2)    \\  \nonumber
       &+& Q(n_2,1) Q(n_1+n_3,2) Q(n_5,1) Q(n_4+n_6,2) + Q(n_1,1) Q(n_2+n_3,2) Q(n_5,1) Q(n_4+n_6,2)    \\  \nonumber
       &+& Q(n_2,1) Q(n_3,1) Q(n_1+n_5,2) Q(n_4+n_6,2) + Q(n_1,1) Q(n_3,1) Q(n_2+n_5,2) Q(n_4+n_6,2)    \\  \nonumber
       &+& Q(n_1,1) Q(n_2,1) Q(n_3+n_5,2) Q(n_4+n_6,2) + Q(n_1+n_2,2) Q(n_3,1) Q(n_4,1) Q(n_5+n_6,2)    \\  \nonumber
       &+& Q(n_2,1) Q(n_1+n_3,2) Q(n_4,1) Q(n_5+n_6,2) + Q(n_1,1) Q(n_2+n_3,2) Q(n_4,1) Q(n_5+n_6,2)    \\  \nonumber
       &+& Q(n_2,1) Q(n_3,1) Q(n_1+n_4,2) Q(n_5+n_6,2) + Q(n_1,1) Q(n_3,1) Q(n_2+n_4,2) Q(n_5+n_6,2)    \\  \nonumber
       &+& Q(n_1,1) Q(n_2,1) Q(n_3+n_4,2) Q(n_5+n_6,2),   \\  \nonumber
\end{eqnarray}
\begin{eqnarray}
 B5 &=& 
  -  2 Q(n_6,1) Q(n_1+n_5,2) Q(n_2+n_3+n_4,3) - 2 Q(n_6,1) Q(n_2+n_5,2) Q(n_1+n_3+n_4,3)    \\  \nonumber
 &-& 2 Q(n_6,1) Q(n_3+n_4,2) Q(n_1+n_2+n_5,3) - 2 Q(n_6,1) Q(n_3+n_5,2) Q(n_1+n_2+n_4,3)    \\  \nonumber
 &-& 2 Q(n_6,1) Q(n_2+n_4,2) Q(n_1+n_3+n_5,3) - 2 Q(n_6,1) Q(n_1+n_4,2) Q(n_2+n_3+n_5,3)    \\  \nonumber
 &-& 2 Q(n_6,1) Q(n_4+n_5,2) Q(n_1+n_2+n_3,3) - 2 Q(n_6,1) Q(n_2+n_3,2) Q(n_1+n_4+n_5,3)    \\  \nonumber
 &-& 2 Q(n_6,1) Q(n_1+n_3,2) Q(n_2+n_4+n_5,3) - 2 Q(n_6,1) Q(n_1+n_2,2) Q(n_3+n_4+n_5,3)    \\  \nonumber
 &-& 2 Q(n_5,1) Q(n_1+n_6,2) Q(n_2+n_3+n_4,3) - 2 Q(n_5,1) Q(n_2+n_6,2) Q(n_1+n_3+n_4,3)    \\  \nonumber
 &-& 2 Q(n_5,1) Q(n_3+n_4,2) Q(n_1+n_2+n_6,3) - 2 Q(n_5,1) Q(n_2+n_4,2) Q(n_1+n_3+n_6,3)    \\  \nonumber
 &-& 2 Q(n_5,1) Q(n_1+n_4,2) Q(n_2+n_3+n_6,3) - 2 Q(n_5,1) Q(n_3+n_6,2) Q(n_1+n_2+n_4,3)    \\  \nonumber
 &-& 2 Q(n_5,1) Q(n_4+n_6,2) Q(n_1+n_2+n_3,3) - 2 Q(n_5,1) Q(n_2+n_3,2) Q(n_1+n_4+n_6,3)    \\  \nonumber
 &-& 2 Q(n_5,1) Q(n_1+n_3,2) Q(n_2+n_4+n_6,3) - 2 Q(n_5,1) Q(n_1+n_2,2) Q(n_3+n_4+n_6,3)    \\  \nonumber
 &-& 2 Q(n_4,1) Q(n_1+n_6,2) Q(n_2+n_3+n_5,3) - 2 Q(n_4,1) Q(n_2+n_6,2) Q(n_1+n_3+n_5,3)    \\  \nonumber
 &-& 2 Q(n_4,1) Q(n_3+n_5,2) Q(n_1+n_2+n_6,3) - 2 Q(n_4,1) Q(n_3+n_6,2) Q(n_1+n_2+n_5,3)    \\  \nonumber
 &-& 2 Q(n_4,1) Q(n_2+n_5,2) Q(n_1+n_3+n_6,3) - 2 Q(n_4,1) Q(n_1+n_5,2) Q(n_2+n_3+n_6,3)    \\  \nonumber
 &-& 2 Q(n_4,1) Q(n_5+n_6,2) Q(n_1+n_2+n_3,3) - 2 Q(n_4,1) Q(n_2+n_3,2) Q(n_1+n_5+n_6,3)    \\  \nonumber
 &-& 2 Q(n_4,1) Q(n_1+n_3,2) Q(n_2+n_5+n_6,3) - 2 Q(n_4,1) Q(n_1+n_2,2) Q(n_3+n_5+n_6,3)    \\  \nonumber
 &-& 2 Q(n_3,1) Q(n_1+n_6,2) Q(n_2+n_4+n_5,3) - 2 Q(n_3,1) Q(n_2+n_6,2) Q(n_1+n_4+n_5,3)    \\  \nonumber
 &-& 2 Q(n_3,1) Q(n_4+n_5,2) Q(n_1+n_2+n_6,3) - 2 Q(n_3,1) Q(n_1+n_2,2) Q(n_4+n_5+n_6,3)    \\  \nonumber
 &-& 2 Q(n_3,1) Q(n_1+n_4,2) Q(n_2+n_5+n_6,3) - 2 Q(n_3,1) Q(n_2+n_4,2) Q(n_1+n_5+n_6,3)    \\  \nonumber
 &-& 2 Q(n_3,1) Q(n_5+n_6,2) Q(n_1+n_2+n_4,3) - 2 Q(n_3,1) Q(n_1+n_5,2) Q(n_2+n_4+n_6,3)    \\  \nonumber
 &-& 2 Q(n_3,1) Q(n_2+n_5,2) Q(n_1+n_4+n_6,3) - 2 Q(n_3,1) Q(n_4+n_6,2) Q(n_1+n_2+n_5,3)    \\  \nonumber
 &-& 2 Q(n_2,1) Q(n_1+n_6,2) Q(n_3+n_4+n_5,3) - 2 Q(n_2,1) Q(n_3+n_6,2) Q(n_1+n_4+n_5,3)    \\  \nonumber
 &-& 2 Q(n_2,1) Q(n_4+n_5,2) Q(n_1+n_3+n_6,3) - 2 Q(n_2,1) Q(n_1+n_3,2) Q(n_4+n_5+n_6,3)    \\  \nonumber
 &-& 2 Q(n_2,1) Q(n_1+n_4,2) Q(n_3+n_5+n_6,3) - 2 Q(n_2,1) Q(n_1+n_5,2) Q(n_3+n_4+n_6,3)    \\  \nonumber
 &-& 2 Q(n_2,1) Q(n_3+n_4,2) Q(n_1+n_5+n_6,3) - 2 Q(n_2,1) Q(n_3+n_5,2) Q(n_1+n_4+n_6,3)    \\  \nonumber
 &-& 2 Q(n_2,1) Q(n_4+n_6,2) Q(n_1+n_3+n_5,3) - 2 Q(n_2,1) Q(n_5+n_6,2) Q(n_1+n_3+n_4,3)    \\  \nonumber
 &-& 2 Q(n_1,1) Q(n_4+n_5,2) Q(n_2+n_3+n_6,3) - 2 Q(n_1,1) Q(n_3+n_6,2) Q(n_2+n_4+n_5,3)    \\  \nonumber
 &-& 2 Q(n_1,1) Q(n_2+n_6,2) Q(n_3+n_4+n_5,3) - 2 Q(n_1,1) Q(n_2+n_3,2) Q(n_4+n_5+n_6,3)    \\  \nonumber
 &-& 2 Q(n_1,1) Q(n_2+n_4,2) Q(n_3+n_5+n_6,3) - 2 Q(n_1,1) Q(n_2+n_5,2) Q(n_3+n_4+n_6,3)    \\  \nonumber
 &-& 2 Q(n_1,1) Q(n_3+n_4,2) Q(n_2+n_5+n_6,3) - 2 Q(n_1,1) Q(n_3+n_5,2) Q(n_2+n_4+n_6,3)    \\  \nonumber
 &-& 2 Q(n_1,1) Q(n_4+n_6,2) Q(n_2+n_3+n_5,3) - 2 Q(n_1,1) Q(n_5+n_6,2) Q(n_2+n_3+n_4,3),   \\  \nonumber
\end{eqnarray}
\begin{eqnarray}
B6 &=&   
 -  6 Q(n_1,1) Q(n_3,1) Q(n_2+n_4+n_5+n_6,4) - 6 Q(n_1,1) Q(n_2,1) Q(n_3+n_4+n_5+n_6,4)    \\  \nonumber
&-& 6 Q(n_1,1) Q(n_4,1) Q(n_2+n_3+n_5+n_6,4) - 6 Q(n_1,1) Q(n_5,1) Q(n_2+n_3+n_4+n_6,4)    \\  \nonumber
&-& 6 Q(n_1,1) Q(n_6,1) Q(n_2+n_3+n_4+n_5,4) - 6 Q(n_2,1) Q(n_4,1) Q(n_1+n_3+n_5+n_6,4)    \\  \nonumber
&-& 6 Q(n_2,1) Q(n_3,1) Q(n_1+n_4+n_5+n_6,4) - 6 Q(n_2,1) Q(n_5,1) Q(n_1+n_3+n_4+n_6,4)    \\  \nonumber
&-& 6 Q(n_2,1) Q(n_6,1) Q(n_1+n_3+n_4+n_5,4) - 6 Q(n_3,1) Q(n_5,1) Q(n_1+n_2+n_4+n_6,4)    \\  \nonumber
&-& 6 Q(n_3,1) Q(n_4,1) Q(n_1+n_2+n_5+n_6,4) - 6 Q(n_3,1) Q(n_6,1) Q(n_1+n_2+n_4+n_5,4)    \\  \nonumber
&-& 6 Q(n_4,1) Q(n_6,1) Q(n_1+n_2+n_3+n_5,4) - 6 Q(n_4,1) Q(n_5,1) Q(n_1+n_2+n_3+n_6,4)    \\  \nonumber
&-& 6 Q(n_5,1) Q(n_6,1) Q(n_1+n_2+n_3+n_4,4),   \\  \nonumber
\end{eqnarray}
\begin{eqnarray}
B7 &=& 
    -  Q(n_1+n_2,2) Q(n_3+n_4,2) Q(n_5+n_6,2) - Q(n_1+n_2,2) Q(n_3+n_5,2) Q(n_4+n_6,2)    \\  \nonumber
   &-& Q(n_1+n_2,2) Q(n_4+n_5,2) Q(n_3+n_6,2) - Q(n_1+n_3,2) Q(n_2+n_4,2) Q(n_5+n_6,2)    \\  \nonumber
   &-& Q(n_1+n_3,2) Q(n_2+n_5,2) Q(n_4+n_6,2) - Q(n_1+n_3,2) Q(n_2+n_6,2) Q(n_4+n_5,2)    \\  \nonumber
   &-& Q(n_1+n_4,2) Q(n_2+n_3,2) Q(n_5+n_6,2) - Q(n_1+n_4,2) Q(n_2+n_5,2) Q(n_3+n_6,2)    \\  \nonumber
   &-& Q(n_1+n_4,2) Q(n_2+n_6,2) Q(n_3+n_5,2) - Q(n_1+n_5,2) Q(n_2+n_3,2) Q(n_4+n_6,2)    \\  \nonumber
   &-& Q(n_1+n_5,2) Q(n_2+n_4,2) Q(n_3+n_6,2) - Q(n_1+n_5,2) Q(n_2+n_6,2) Q(n_3+n_4,2)    \\  \nonumber
   &-& Q(n_1+n_6,2) Q(n_2+n_3,2) Q(n_4+n_5,2) - Q(n_1+n_6,2) Q(n_2+n_4,2) Q(n_3+n_5,2)    \\  \nonumber
   &-& Q(n_1+n_6,2) Q(n_2+n_5,2) Q(n_3+n_4,2),    \\  \nonumber
\end{eqnarray}
\begin{eqnarray}
B8 &=& 
       4 Q(n_1+n_2+n_3,3) Q(n_4+n_5+n_6,3) + 4 Q(n_1+n_2+n_4,3) Q(n_3+n_5+n_6,3)   \\  \nonumber
   &+& 4 Q(n_1+n_2+n_5,3) Q(n_3+n_4+n_6,3) + 4 Q(n_1+n_3+n_4,3) Q(n_2+n_5+n_6,3)   \\  \nonumber
   &+& 4 Q(n_1+n_3+n_5,3) Q(n_2+n_4+n_6,3) + 4 Q(n_1+n_4+n_5,3) Q(n_2+n_3+n_6,3)   \\  \nonumber
   &+& 4 Q(n_2+n_3+n_4,3) Q(n_1+n_5+n_6,3) + 4 Q(n_2+n_3+n_5,3) Q(n_1+n_4+n_6,3)   \\  \nonumber
   &+& 4 Q(n_2+n_4+n_5,3) Q(n_1+n_3+n_6,3) + 4 Q(n_3+n_4+n_5,3) Q(n_1+n_2+n_6,3),  \\  \nonumber
\end{eqnarray}
\begin{eqnarray}
B9 &=& 
       6 Q(n_1+n_2,2) Q(n_3+n_4+n_5+n_6,4) + 6 Q(n_1+n_3,2) Q(n_2+n_4+n_5+n_6,4)   \\  \nonumber
   &+& 6 Q(n_1+n_4,2) Q(n_2+n_3+n_5+n_6,4) + 6 Q(n_1+n_5,2) Q(n_2+n_3+n_4+n_6,4)   \\  \nonumber
   &+& 6 Q(n_1+n_6,2) Q(n_2+n_3+n_4+n_5,4) + 6 Q(n_2+n_3,2) Q(n_1+n_4+n_5+n_6,4)   \\  \nonumber
   &+& 6 Q(n_2+n_4,2) Q(n_1+n_3+n_5+n_6,4) + 6 Q(n_2+n_5,2) Q(n_1+n_3+n_4+n_6,4)   \\  \nonumber
   &+& 6 Q(n_2+n_6,2) Q(n_1+n_3+n_4+n_5,4) + 6 Q(n_3+n_4,2) Q(n_1+n_2+n_5+n_6,4)   \\  \nonumber
   &+& 6 Q(n_3+n_5,2) Q(n_1+n_2+n_4+n_6,4) + 6 Q(n_3+n_6,2) Q(n_1+n_2+n_4+n_5,4)   \\  \nonumber
   &+& 6 Q(n_4+n_5,2) Q(n_1+n_2+n_3+n_6,4) + 6 Q(n_4+n_6,2) Q(n_1+n_2+n_3+n_5,4)   \\  \nonumber
   &+& 6 Q(n_5+n_6,2) Q(n_1+n_2+n_3+n_4,4),   \\  \nonumber
\end{eqnarray}
\begin{eqnarray}
B10 &=& 
        24 Q(n_1,1) Q(n_2+n_3+n_4+n_5+n_6,5) + 24 Q(n_2,1) Q(n_1+n_3+n_4+n_5+n_6,5)   \\  \nonumber
    &+& 24 Q(n_3,1) Q(n_1+n_2+n_4+n_5+n_6,5) + 24 Q(n_4,1) Q(n_1+n_2+n_3+n_5+n_6,5)   \\  \nonumber
    &+& 24 Q(n_5,1) Q(n_1+n_2+n_3+n_4+n_6,5) + 24 Q(n_6,1) Q(n_1+n_2+n_3+n_4+n_5,5),  \\  \nonumber
\end{eqnarray}
\begin{eqnarray}
B11 &=& -120 Q(n_1+n_2+n_3+n_4+n_5+n_6,6).   \\  \nonumber
\end{eqnarray}
}

Also the two- and multi-particle correlations are given via the subevents cumulant methods~\cite{Jia:2017hbm,Huo:2017nms,Zhang:2018lls,Magdy:2020bhd}. The observables discussed can be given in terms of the flow vectors $Q_{n}$ as;
\begin{eqnarray}\label{eq:B-1}
Q(Sub,n,k)              &=&  \sum^{M}_{i=1} \omega^{k}_{i} e^{in\varphi^{Sub}_{i}},
\end{eqnarray}
where $\phi_i$ is the azimuthal angle of the $\mathit{i}^{th}$ particle. The order $n$ can take positive or negative values, for $n$=0 Eq.~\ref{eq:A2-1} $Q(Sub,0,k)$ gives the event multiplicity and $Sub$ refer to the subevent used. In this work the two subevents used with  $\Delta\eta~=~\eta_{1}-\eta_{2}~ > 0.7$  between the subevents $\textit{A}$ and $\textit{B}$ (\textit{i.e.}, $\eta_{X1}~ > 0.35$ and $\eta_{X2}~ < -0.35$). 
\begin{itemize}
\item{The two-subevents two-particle $SC(n_1,n_2)$;}
\end{itemize}

\begin{eqnarray}\label{eq:B-2}
SC(n_1,n_2)    &=& \langle Two_S(n_1,n_2) \rangle / \langle Two_S(0,0) \rangle, \\  \nonumber
Two_S(n_1,n_2) &=&  QS(X1,n_1,1) QS(X2,n_2,1).
\end{eqnarray}

\begin{itemize}
\item{The two-subevents three-particle $ASC(n_1,n_2,n_3)$;}
\end{itemize}

\begin{eqnarray}\label{eq:A2-1}
ASC(n_1,n_2,n_3)     &=& \langle Three_S(n_1,n_2,n_3) \rangle / \langle Three_S(0,0,0) \rangle, \\  \nonumber
Three_S(n_1,n_2,n_3) &=&  ( QS(X1,n_1,1) QS(X1,n_2,1) - QS(X1,n_1+n_2,2) )  QS(X2,n_3,1).
\end{eqnarray}

\begin{itemize}
\item{The two-subevents four-particle $SC(n_1,n_2,n_3,n_4)$ and $ASC(n_1,n_2,n_3,n_4)$;}
\end{itemize}

\begin{eqnarray}\label{eq:A2-1}
ASC(n_1,n_2,n_3,n_4)    &=& \langle Four_S(n_1,n_2,n_3,n_4) \rangle / \langle Four_S(0,0,0,0) \rangle \\  
Four_S(n_1,n_2,n_3,n_4) &=& (QS(X1,n_1,1) QS(X1,n_2,1) - QS(X1,n_1+n_2,2)) \\  \nonumber
                    & & (QS(X2,n_3,1) QS(X2,n_4,1) - QS(X2,n_3+n_4,2)).                  
\end{eqnarray}

\end{widetext}

\bibliography{ref} 
\end{document}